\documentclass[a4paper,fleqn,usenatbib]{mnras}

\usepackage{graphicx}	
\usepackage{amsmath}	
\usepackage{amssymb}	
\usepackage{comment}

\newcommand{\redmap}{redMaPPer}

\defcitealias{Baxter:2015}{B15}
\defcitealias{George:2015}{G15}
\defcitealias{Melchior:2017}{M17}
\defcitealias{Simet:2017}{S17}

\title[DES+SPT CMB Cluster Lensing]{A Measurement of CMB Cluster
  Lensing with SPT and DES Year 1 Data}

\author[Baxter et al.]{
\normalsize
\parbox{\textwidth}{
E.~J.~Baxter$^{1}$\thanks{E-Mail: ebax@sas.upenn.edu},
S.~Raghunathan$^{2}$,
T.~M.~Crawford$^{3,4}$,
P.~Fosalba$^{5}$,
Z.~Hou$^{3,4}$,
G.~P.~Holder$^{6,7,8}$,
Y.~Omori$^{9}$,
S.~Patil$^{2}$,
E.~Rozo$^{10}$,
T. M. C.~Abbott$^{11}$
% ...
}
\vspace{0.4cm} \\
\parbox{\textwidth}{Author affiliations are listed at the end of this paper}}

\date{Last updated \today}

\pubyear{2017}

\begin{document}
\label{firstpage}
\pagerange{\pageref{firstpage}--\pageref{lastpage}}
\maketitle


\begin{abstract}
Clusters of galaxies gravitationally lens the cosmic microwave
background (CMB) radiation, resulting in a distinct imprint in the CMB
on arcminute scales.  Measurement of this effect offers a promising
way to constrain the masses of galaxy clusters, particularly those at
high redshift.  We use CMB maps from the South Pole Telescope
Sunyaev-Zel'dovich (SZ) survey to measure the CMB lensing signal
around galaxy clusters identified in optical imaging from first year
observations of the Dark Energy Survey.  The cluster catalog used in
this analysis contains 3697 members with mean redshift of $\bar{z} =
0.45$.  We detect lensing of the CMB by the galaxy clusters at
$8.1\sigma$ significance.  Using the measured lensing signal, we
constrain the amplitude of the relation between cluster mass and
optical richness to roughly $17\%$ precision, finding good agreement
with recent constraints obtained with galaxy lensing.  The error
budget is dominated by statistical noise but includes significant
contributions from systematic biases due to the thermal SZ effect and
cluster miscentering.
\end{abstract}




\section{Introduction}
\label{sec:introduction}

Cosmic microwave background (CMB) photons passing near massive galaxy
clusters are gravitationally deflected, leading to small-amplitude
(typically $\lesssim 10 \,\mu {\rm K}$) distortions in the observed
CMB on arcminute scales.  As pointed out by several authors
\citep[e.g.][]{Seljak:2000, Holder:2004, Vale:2004, Dodelson:2004,
  Lewis:2006, Hu:2007}, these distortions can be used to measure the
masses of galaxy clusters.  Because gravitational lensing is sensitive
to the total cluster mass, cluster masses determined from the CMB
lensing signal are in principle robust to uncertainties on baryonic
processes ocurring inside the clusters.  In contrast, cluster
observables such as the thermal Sunyaev-Zel'dovich (tSZ) decrement,
X-ray temperature, and cluster richness may depend on complicated
baryonic physics that can introduce systematic uncertainty into the
cluster mass-observable relations.  Systematic uncertainty on cluster
masses dominates the error budget of most recent cluster abundance
constraints on cosmology \citep[e.g.][]{Planck2015:XXIV, deHaan:2016,
  Rozo:2010, Mantz:2015}.

Cluster masses can also be inferred from gravitationally induced
shearing of images of background galaxies \citep[for a review see
][]{Hoekstra:2013}.  However, at high redshift, measurements of
cluster masses with galaxy lensing become challenging because the
source galaxies become harder to detect and their shapes and redshifts
become more difficult to measure \citep[e.g.][]{Hoekstra:2001}.  CMB
cluster lensing, on the other hand, has the advantage that the signal
to noise is roughly constant with cluster redshift \citep{Lewis:2006}.
Furthermore, CMB cluster lensing is not sensitive to many of the
sources of systematic error that affect estimates of cluster mass
derived from galaxy shear, including PSF modeling errors
\citep[e.g.][]{Jarvis:2016}, biases in the photometric redshift
estimates of source galaxies \citep[e.g.][]{Melchior:2017}, and
contamination of the shear sample with unlensed cluster galaxies
\citep[e.g.][]{Applegate:2014, Melchior:2017}.  Consequently, even at
low redshifts where CMB lensing-derived constraints on cluster masses
may not be statistically competitive with galaxy lensing-derived
constraints, CMB cluster lensing offers an important test of
systematic errors associated with galaxy lensing.

The CMB lensing signal induced by galaxy clusters was first measured
around clusters detected in the South Pole Telescope (SPT)
Sunyaev-Zel'dovich (SZ) Survey by \citet{Baxter:2015} (henceforth
\citetalias{Baxter:2015}).  A similar measurement around
Planck-detected clusters was performed in \citet{PlanckXXIV:2015}.
Both of these early measurements used the CMB cluster lensing signal
to place (weak) constraints on the scaling between the lensing-derived
mass and the mass inferred from measurement of the tSZ.  Related work
by \citet{Madhavacheril:2015} used CMB lensing to constrain the mean
mass of optically selected CMASS galaxies \citep{Eisenstein:2011,
  Dawson:2013,Ahn:2014} using CMB data from the Atacama Cosmology
Telescope Polarimeter (ACTPol).  Recently, \citet{Geach:2017} used CMB
lensing measurements derived from Planck data to constrain the masses
of clusters detected in the Sloan Digital Sky Survey.

The aim of this work is to measure the CMB cluster lensing signal
around galaxy clusters identified in optical imaging from year one
(Y1) Dark Energy Survey (DES) observations and to use the measurement
of CMB cluster lensing to calibrate the relation between cluster
mass and optical richness.  To this end, we employ the same SPT-SZ CMB
temperature maps as used in \citetalias{Baxter:2015}.  However, the
galaxy cluster sample employed here is significantly expanded relative
to that used in \citetalias{Baxter:2015}.  \citetalias{Baxter:2015}
measured the CMB cluster lensing signal using 513 SZ-selected
clusters; here we use 3697 clusters identified using the \redmap{}
\citep{Rykoff:2014} algorithm applied to DES imaging.

This work also represents a significant departure in methodology from
the \citetalias{Baxter:2015} analysis.  \citetalias{Baxter:2015}
defined a map-space likelihood for the observed CMB temperature
measurements around a cluster as a function of cluster mass, and used
that likelihood to constrain the stacked cluster mass of the sample.
In this work, we employ the more standard quadratic estimator approach
to estimate the lensing convergence, $\kappa$, in small cutouts of the
CMB around the galaxy clusters.  The primary advantage of the
quadratic estimator approach employed here is its robustness to
important sources of systematic error.  With minor modification to the
standard filters used to construct the quadratic estimator, we find
that the estimator is fairly robust to the presence of tSZ signal
around the cluster.  Additionally, the quadratic estimator is less
sensitive to other sources of systematic error, such as foreground
lensing.  Consequently, in this analysis we are able to directly use
the low-noise 150~GHz CMB maps from the SPT rather than creating
higher noise tSZ-free maps from multi-frequency data.

We fit the stacked CMB lensing-derived $\kappa$ measurements around
the \redmap{} clusters to place constraints on the \redmap{}
mass-richness relation.  We show that two important sources of
systematic error for these constraints are cluster miscentering and
contamination of the lensing estimator by tSZ signal.

The structure of the paper is as follows: in
\S\ref{sec:quad_estimator} we introduce the formalism for computing
$\kappa$ from CMB temperature maps around clusters; \S\ref{sec:data}
describes the SPT and DES datasets used in this work;
\S\ref{sec:measurement} describes the pipeline we have developed for
measuring CMB lensing around galaxy clusters; \S\ref{sec:modelfit}
describes the process of fitting the lensing measurements to obtain
constraints on the masses of the clusters in our sample;
\S\ref{sec:simulations} describes the simulations we have developed to
test the analysis pipeline.  Our results are described in
\S\ref{sec:results} and discussion of these results is presented in
\S\ref{sec:discussion}.

Throughout this analysis we assume the best-fit $\Lambda$CDM
cosmological model from the Planck TT,TE,EE+lowP fits in
\citet{Planck:2016}.  Cluster masses are described in terms of
$M_{200m}$, the mass enclosed within a sphere of radius $R_{200m}$
centered on the cluster.  $R_{200m}$ is in turn the distance from the
cluster center at which the mean enclosed density is 200 times
the mean density of the Universe at the redshift of the cluster.

\section{A Quadratic Estimator for $\kappa$}
\label{sec:quad_estimator}

Gravitational lensing of the CMB remaps the image of the last
scattering surface.  The observed temperature in direction $\hat{n}$
is equal to the unlensed temperature shifted by the deflection angle,
$\nabla \phi$:
\begin{eqnarray}
T(\hat{n}) = \bar{T}(\hat{n} + \nabla \phi),
\end{eqnarray}
where the overbar is used to indicate unlensed quantities, and $\phi$
is the lensing potential.  The lensing potential is in turn related to
the convergence, $\kappa$, via
\begin{eqnarray}
\kappa = -\frac{1}{2} \nabla^2 \phi.
\end{eqnarray}
As a result of diffusion damping, the primordial CMB has little power
on arcminute scales.  Because the deflections induced by cluster
lensing are at most a few arcminutes we can approximate the CMB as a
pure gradient over the scales at which the deflections are occurring.
This allows us to re-write the temperature of the lensed CMB as
\begin{eqnarray}
\label{eq:grad_approx}
T \approx \bar{T} + \nabla \bar{T} \cdot \nabla \phi,
\end{eqnarray}
where we have suppressed the dependence on $\hat{n}$ for clarity.
Eq.~\ref{eq:grad_approx} makes it apparent that lensing introduces a
correlation between fluctuations in CMB temperature field and the
background gradient field.  The quadratic estimator introduced by
\citet{Hu:2002} recovers an estimate of $\kappa$ by identifying these
correlations in the maps: a filtered gradient map is multiplied by a
high-pass filtered temperature map (see below).  This estimator is
quadratic in the sense that it involves two powers of the temperature
field.  \citet{Hu:2002} showed how to construct Fourier-space filters
that give the minimum variance estimate of $\kappa$ in this approach.

As pointed out by \citet{Hu:2007} and others, the quadratic estimator
formulated by \citet{Hu:2002} is biased in regions of high $\kappa$.
This bias can be understood as resulting from the fact that a very
massive lens (such as a cluster) will magnify the CMB gradient behind
it, effectively decreasing the magnitude of the estimated gradient
field; the result is that the $\kappa$ estimate is biased low.  To
remove this bias, \citet{Hu:2007} showed that one could simply apply
an additional low pass filter to the maps before estimating the
gradient field.  This additional filter separates out the scales used
to estimate the gradient from those used to measure the small scale
CMB fluctuations caused by lensing, and thereby removes the bias.  The
filter scale used, $l_G$, becomes an additional parameter of the
analysis, but the results are not expected to be very sensitive to
variations in this scale.

Explicitly, the \citet{Hu:2007} estimator for $\kappa$ is
\begin{eqnarray}
\label{eq:qe}
\frac{\hat{\kappa}_{\vec{l}}}{A_{\vec{l}}} = - \int d^2 \hat{n} \, e^{-i \hat{n} \cdot \vec{l}} \mathrm{Re} \left\{ \nabla \cdot \left[G(\hat{n}) L(\hat{n}) \right] \right\},
\end{eqnarray}
where $G(\hat{n})$ is the filtered gradient field, $L(\hat{n})$ is the
filtered temperature field, and $A_{\vec{l}}$ is a normalization term.
The filered gradient field can be written in Fourier space as
\begin{eqnarray}
G(\vec{l}) = i \vec{l} W_l T_{\vec{l}},
\end{eqnarray}
where $T_{\vec{l}}$ is the Fourier transform of the temperature field
and $W_l$ represents the filter function.  Following \citet{Hu:2007}, we set 
\begin{eqnarray}
W_l = 
  \begin{cases}
    \bar{C}_l\left( C_l + N_l \right)^{-1} & \text{if $l \leq l_G$} \\
    0 & \text{if $l > l_G$},
  \end{cases}
\end{eqnarray}
where $C_l$ ($\bar{C}_l$) is the (un)lensed temperature power
spectrum, $N(\vec{l})$ is the noise power spectrum, and $l_G$ is the
parameter that controls the gradient filter scale.  The filtered
temperature map is generated using inverse-variance weighting:
\begin{eqnarray}
L(\vec{l}) = \overline{W}_l T_{\vec{l}},
\end{eqnarray}
with $\overline{W}_l = (C_l + N_l)^{-1}$.

In order to return an unbiased estimate of $\kappa$, the normalization
factor $A_{\vec{l}}$ introduced in Eq.~\ref{eq:qe} should be
\begin{eqnarray}
\frac{1}{A_{\vec{l}}} = \frac{2}{l^2} \int \frac{dl_1}{(2\pi)^2} \vec{l}\cdot\vec{l}_1 W_{\vec{l}_1} \overline{W}_{\vec{l}_2} f^{TT}(\vec{l}_1, \vec{l}_2),
\end{eqnarray}
where
\begin{eqnarray}
f^{TT}(\vec{l}_1, \vec{l}_2) = (\vec{l} \cdot \vec{l}_1) \bar{C}_{l_1} + (\vec{l} \cdot \vec{l}_2) \bar{C}_{l_2},
\end{eqnarray}
and $\vec{l} = \vec{l}_1 + \vec{l}_2$.

In the analysis presented below, we will apply the \citet{Hu:2007}
quadratic estimator to estimate $\kappa$ in cutouts of the SPT CMB
maps around galaxy clusters.  Because $\kappa$ is directly related to
the integrated mass along the line of sight, by fitting a model to the
recovered $\kappa$ we can extract constraints on the masses of the
clusters in our sample.

\section{Data}
\label{sec:data}

\subsection{CMB maps from SPT}
\label{sec:cmb_data}

The SPT is a 10-meter millimeter/submillimeter telescope operating at
the geographical South Pole \citep{Carlstrom:2011}.  The CMB maps used
in this analysis are from the 2500 sq. deg. SPT-SZ survey, which
mapped the sky in three frequency bands centered at 95, 150 and
220~GHz over an observation period from 2008 to 2011
\citep{Story:2013}.  We use only the SPT 150~GHz maps in this analysis
as these have the lowest noise.

SPT observations are divided into patches of the sky ({\it fields})
that each have an area $\gtrsim 100$ sq. deg.  For most fields, SPT
scans the sky in strips of constant elevation, first moving left, then
right, followed by a step in elevation.  For one field (ra21hdec-50),
a modified scanning strategy was used for some observations, but we
use only the azimuthal scan data from that field in this analysis.
The time ordered data from these scans is filtered to prevent aliasing
of high-frequency noise and to remove atmospheric and instrumental
noise.  The time ordered data is processed into maps with 0.5
arcminute resolution using the Lambert equal-area azimuthal
projection.  The maps used here are identical to those used in the
\citet{George:2015} (hereafter \citetalias{George:2015}) analysis, and
we refer readers to that work for more details of the map making
process.

The signal on the sky observed with the SPT is modified by a response
function consisting of a beam function and a transfer function.  The
beam function describes the smearing of sky sources as a result of the
finite aperture of the SPT primary mirror.  The transfer function
accounts for time-domain filtering applied to the SPT signal.  The
total SPT response function to a mode $\vec{l}$ on the sky can be
modeled as the product $B(l)T(\vec{l})$, where $B(l)$ and $T(\vec{l})$
are the beam and transfer functions, respectively, and $B(l)$ only
depends on $l = |\vec{l}|$ since the beam is close to rotationally
invariant.

The amount of time spent observing each field in the SPT-SZ survey is
not constant, and the effective depth across a field varies slightly
as a result of scanning strategy.  To characterize the varying depth
levels between fields and within a field, we define the weight,
$\omega_i$, at map pixel $i$.  The weight is roughly proportional to
the inverse variance of the map noise at that pixel.  We will use maps
of the weight across the sky to calculate the appropriate noise power
spectrum for each cluster cutout.

Before computing $\kappa$ from the CMB maps, point sources detected in
the maps at $5\sigma$ are masked and inpainted with Gaussian noise
that matches the noise level of the SPT maps.  The point source
catalog used for this purpose is the same as used in
\citet{George:2015} for masking.

\subsection{\redmap{} cluster catalog from DES}
\label{sec:cluster_catalog}

DES is a five-year optical imaging survey of 5000 sq. deg. of the
southern sky using the Dark Energy Camera on the Blanco Telescope
\citep{Flaugher2015}.  In this analysis, we make use of first year
(Y1) DES data, which covers roughly 1800 sq. deg. of sky
\citep{Diehl:2014, Drlica-Wagner:2017}.  The total area of overlap
between Y1 observations and the SPT-SZ survey is roughly 1500 sq. deg.

Galaxy clusters were identified in the Y1 data using the \redmap{}
algorithm \citep{Rykoff:2014}.  The application of \redmap{} to early
DES Science Verification (SV) data is described in
\citet{Rykoff:2016}.  The application of \redmap{} to Y1 data will be
described in more detail in an upcoming publication (McClintock et
al., in prep.).  \redmap{} identifies cluster candidates as
overdensities of red-sequence galaxies on the sky.  For each cluster
candidate, \redmap{} determines a list of possible cluster member
galaxies and their corresponding membership probabilities.  The
\redmap{} estimate of the cluster richness, $\lambda$, is defined as
the sum over these membership probabilities for each
cluster. \redmap{} also computes centering probabilities, $P_{\rm
  cen}$, which characterize the probability that a member galaxy is at
the center of the cluster.  We treat the galaxy with the highest
$P_{\rm cen}$ as the cluster center in this analysis, but also
consider the effects of various degrees of miscentering.  We will also
use the \redmap{} estimates of the cluster redshifts in this analysis;
these are expected to be accurate to roughly $\sigma_z \sim
0.01(1+z)$.

We consider only the volume-limited \redmap{} DES Y1 catalog,
restricted to clusters with richness $\lambda > 20$, resulting in a
total of 7066 clusters.  We further impose the requirements that the
minimum SPT-defined weight in a $\sim 2^{\circ} \times 2^{\circ}$
cutout around the cluster (see \S\ref{sec:measurement} for a more
detailed description of the cutouts) is greater than zero and is at
least 80\% of the weight at the cluster location.  These restrictions
ensure that we do not include clusters outside of the SPT fields or
clusters for which the weight is varying significantly across the
cutout (as may occur near the field boundaries).  After imposing this
restriction, the cluster catalog is reduced to 4552 clusters.
Finally, as will be described in more detail in \S\ref{sec:tsz}, we
find that the presence of tSZ signal around high-mass clusters can
bias the $\kappa$ reconstruction.  Because the amplitude of the tSZ
signal scales as $M^{5/3}$, by restricting our analysis to lower mass
clusters we find that we can reduce the tSZ bias to acceptable levels
while preserving much of the lensing signal.  To this end, we employ a
somewhat conservative richness cut, restricting our analysis to
$\lambda < 40$.  This richness threshold corresponds to a mass of
about $3.2\times10^{14} M_{\odot}$ assuming the mass-richness relation
of \citet{Melchior:2017} (hereafter \citetalias{Melchior:2017}).  We
discuss the motivation for the richness cut and tests of potential tSZ
biases in more detail in \S\ref{sec:systematics}.  Imposing the
richness cut yields a final catalog of 3697 clusters ranging in
redshift from roughly $z \sim 0.1$ to $0.7$, with mean redshift
$\bar{z} = 0.45$.

\section{Measurement of $\kappa$}
\label{sec:measurement}

For each DES-identified cluster, we estimate $\kappa$ using
cluster-centered cutouts from the SPT 150~GHz temperature maps
presented in \citetalias{George:2015}.  The CMB temperature maps are
pixelized at 0.5 arcminute resolution and the cutouts are 256 pixels
on a side.

We rotate each cutout so that it is aligned along lines of constant
azimuth and elevation (see e.g. \citealt{Schaffer:2011} for description
of the rotation angles corresponding to the Lambert equal-area
azimuthal projection).  Aligning the cutouts with altitude and azimuth
ensures that the transfer function is the same for every cutout,
significantly simplifying the subsequent analysis.  Rotation is
performed using third order spline interpolation.  The rotated cutouts
are then apodized using a Tukey window with $\alpha = 0.1$.  Our
analysis of simulated data in \S\ref{sec:simulations} confirms that
this choice of apodization is reasonable.

The SPT beam and transfer functions are deconvolved from the cutouts
before the application of the quadratic estimator.  Estimates of the
beam functions used for this purpose are described in
\citet{Story:2013}.  The transfer function deconvolved from each
cutout is an analytic approximation to the true transfer function
consisting of three pieces:
\begin{eqnarray}
\label{eq:tf_model}
T(\vec{l}) = F_1(\vec{l}) F_2(\vec{l}) F_3(\vec{l}),
\end{eqnarray}
where
\begin{eqnarray}
F_1(\vec{l}) &=& \exp(-(l_1/l)^6) \\
F_2(\vec{l}) &=& \exp(-(l_2/l_x)^6) \\
F_3(\vec{l}) &=& \exp(-(l_x/l_3)^6),
\end{eqnarray}
with $l_1 = 500$, $l_2 = 400$, and $l_3 = 15000$.  We have checked that there is negligible difference between the deconvolved cutouts obtained using the above transfer function approximation and the deconvolved cutouts obtained using a more accurate estimate of the transfer function that was computed by mock observing a $\delta$-function signal located at the center of the field.  To prevent highly
filtered modes from introducing numerical problems as a result of
deconvolution, we additionally filter each cutout to remove modes with
$l < l_1$, $l_x < l_2$ and $l_x > l_3$.

\subsection{Noise and Foregrounds}
\label{sec:noise_estimation}

Estimating $\kappa$ near each cluster requires an estimate of the
noise power spectrum in each CMB temperature cutout.  We consider as
noise any contribution to the cutouts that is not CMB and that is not
expected to be correlated with the positions of the galaxy clusters.
Non-CMB signal that is correlated with the clusters --- such as the
cluster tSZ signal --- is treated as a source of systematic error and
is discussed in \S\ref{sec:systematics}.  Each cutout receives
contributions from astronomical, atmospheric, and instrumental noise
sources.  We will take a model-based approach to estimating the
contributions from astronomical noise sources; we estimate the
contribution from atmospheric and instrumental noise directly from the
data.

We first consider the contribution to the cutouts from astronomical
noise sources; such noise is constant over the timescale of the
observations.  In addition to the CMB, the sky signal at 150~GHz
receives significant contributions from several sources, including
galaxies that are bright at microwave frequencies and unresolved tSZ
signal.  The relative contributions of these various sources is
$l$-dependent: at low multipoles ($\ell \lesssim 3000$), the CMB
dominates while at higher multipoles ($\ell \gtrsim 3000$), foreground
emission becomes dominant.  In general, the astrophysical foreground
sources can be approximated as Gaussian random noise.  Although
foreground emission from extragalactic sources such as dusty galaxies
and the tSZ are known to be non-Gaussian at some level
\citep[e.g.][]{Crawford:2014}, the Gaussian approximation should be
sufficient for the noise levels considered here
\citep{vanEngelen:2014}.  The SPT CMB maps also receive some
contribution from galactic foregrounds, such as dust.  However, this
foreground contribution is expected to be significantly below the
contributions from the other foregrounds mentioned
above \citep[e.g.][]{Keisler:2011}.

Once bright point sources have been removed, the dominant foreground
contribution to the sky at 150~GHz comes from dusty, star-forming
galaxies (DSFGs).  The power spectrum of DSFG emission can be divided
into two components: one arising from sources on the sky that are
unclustered, and another arising from sources that are clustered on
the sky.  The unclustered component has an angular power spectrum
given by $C_{l} = C_0$, where $C_0$ is a constant.  Expressed in terms
of $D_{l} = l(l+1)C_l/(2\pi)$, the analysis of
\citetalias{George:2015} found $D_{3000} = 9.16 \pm 0.36 \,\mu{\rm
  K}^2$ for the unclustered component.  The clustered DSFG component, on
the other hand, can be modeled with $D_{l} \propto l^{0.8}$ for $l >
1500$.  The \citetalias{George:2015} analysis found $D_{3000} =
3.46\pm 0.54 \, \mu{\rm K}^2$ for this component.  For $l < 1500$, the
contribution of the clustered DSFG foreground can be ignored.

Two other significant foreground contributors at 150~GHz are radio
galaxy emission below the detection threshold of SPT maps and the tSZ
signal from undetected galaxy groups or clusters. Following
\citetalias{George:2015}, we model the signal from radio galaxies
below the SPT detection threshold as an unclustered component with
$D_{3000} = 1.28 \mu {\rm K}^2$.  To model the tSZ signal from sources
below the detection threshold we use the templates from
\citet{Shaw:2010}, normalized using the constraint from
\citetalias{George:2015}.  \citetalias{George:2015} constrained the
amplitude of the tSZ power spectrum after masking sources above the
detection threshold to be $D_{3000} = 2.33^{+0.8}_{-1.4} \mu{\rm
  K}^2$, and we use this value here.

For our fiducial analysis we make the simplifying approximation that
all emission from astrophysical foregrounds is unlensed by the galaxy
clusters.  In reality, this approximation is not very good.  The
cosmic infrared background (CIB), for instance, is expected to receive
significant contribution from redshifts $1 < z < 3$
\citep[e.g.][]{Bethermin:2013}.  Because the clusters used in this
analysis have $\bar{z} = 0.45$, a significant portion of the CIB may
be lensed by the clusters. However, the precise redshift distribution
of the DSFGs and other foreground sources is not known, making
modeling of foreground lensing difficult.  While treating the
foregrounds as unlensed is incorrect at some level, we will show in
\S\ref{sec:foreground_lensing_systematic} that this simplifying
assumption has little effect on our results.

Unlike astrophysical foregrounds, the contribution from atmospheric
and instrumental noise sources is not constant over the SPT observation
time, allowing us to estimate the contributions to the noise power
spectrum from these sources by differencing maps constructed from
observations taken at different times.  As described in
\S\ref{sec:cmb_data}, SPT fields are observed by scanning the
telescope left and right across the full field at a series of discrete
elevations.  We can form a signal-free map as the combination
$\vec{m}^{\rm diff} = (\vec{m}_L - \vec{m}_R)/2$, where $\vec{m}_L$
and $\vec{m}_R$ are the maps formed from left and right-going scans,
respectively.  Because atmospheric and instrumental noise vary on time
scales that are much shorter than the time difference between the
$\vec{m}_L$ and $\vec{m}_R$ observations, $\vec{m}^{\rm diff}$ should
provide a realization of the instrumental and atmospheric noise.

The SPT-SZ survey spent different amounts of time observing each
field, resulting in field-to-field variations in the effective noise
levels of the resultant CMB maps.  Additional variation in the noise
levels between and within fields occurs as a result of sky projection.
To account for field-to-field variation in the noise level, each
cluster is analyzed using the difference maps for the field in which
it was observed.  To account for variation in the noise level across
the field, we use the SPT weight maps, $\vec{\omega}$.

We first construct a scaled difference map that has effectively
uniform weight by multiplying the map by $\sqrt{\vec{\omega}/\langle
  \omega \rangle}$, where $\langle \omega \rangle$ is the mean weight
across the inner $[0.3, 0.7]$ of the map.  We then compute the
instrumental and atmospheric noise power spectrum at the mean weight
using this scaled difference map.  To determine the estimate of the
instrumental and atmospheric noise power spectrum for the $i$th cutout
we then rescale the noise power spectrum estimate for the scaled
difference map by $\langle \omega \rangle/\omega_i$, where $\omega_i$
is the mean weight across the $i$th cutout.

Finally, because we deconvolve the beam and transfer functions from
the cutouts before applying the quadratic estimator, we must account
for this in the noise power spectrum estimate.  The total noise power
spectrum estimate for the $i$th cutout is then
\begin{eqnarray}
\label{eq:noise_beamdeconv}
N_i(\vec{l}) = N_F(\vec{l}) + \frac{\langle \omega \rangle}{\omega_i}\frac{N_{\rm IA}(\vec{l})}{\left[ B(\vec{l})T(\vec{l}) \right]^2},
\end{eqnarray}
where $N_F(\vec{l})$ is the estimate of the noise contribution from
the astrophysical foregrounds described above, $N_{\rm IA}$ is the
estimate of the instrumental and atmospheric noise from the scaled
difference map, and $B(\vec{l})$ and $T(\vec{l})$ are the beam and
transfer function estimates, respectively.  The estimate of the
foreground noise contribution has no beam or transfer function by
construction, so it does not require the beam or transfer function to
be deconvolved.

\subsection{Stacked, Filtered $\kappa$ estimate}
\label{sec:filtered_kappa}

Given the beam and transfer function deconvolved cutouts and the
estimate of the noise power spectrum in the cutout, we compute
$\kappa$ across the cutouts as described in
\S\ref{sec:quad_estimator}.  When generating the $\kappa$ estimate for
each cutout we use a gradient filter scale of $l_G = 1500$.  In
principle, higher signal to noise could be achieved by setting $l_G =
2000$ as originally suggested by \citet{Hu:2007}.  However, we have
found in tests on simulated data (see \S\ref{sec:tsz}) that the tSZ
signal from massive clusters can lead to a significant bias in the
recovered mass when using $l_G = 2000$.  By using the lower value of
$l_G = 1500$, we find that this bias can be significantly reduced
without significantly degrading the signal to noise.  The low pass
filter removes some of the highly localized tSZ signal while
preserving most of the information about the large scale gradient in
the CMB near the cluster.

By deconvolving the beam function from the cutouts, we increase the
effective noise of small scale modes that are heavily filtered by the
beam (e.g. Eq.~\ref{eq:noise_beamdeconv}).  Such small scale noise is
problematic for our analysis since we attempt to fit the $\kappa$
profiles of the clusters in real-space.  In real-space the filtered
scales are not cleanly separated from the unfiltered scales, and the
resultant small-scale noise introduces numerical problems.  To reduce
the effects of such noise, we filter the $\kappa$ cutouts with a low
pass filter to remove modes with $l < \sqrt{8 \ln 2}/\theta_{FWHM}$,
where $\theta_{FWHM} = 1.3'$ is chosen to roughly match the beam size
of the SPT.

Because the estimate of $\kappa$ at the cluster location depends on
the gradient of the CMB temperature field, there is useful information
for constraining $\kappa$ in the temperature maps at scales well
beyond the halo virial radius.  However, once the estimate of $\kappa$
has been computed, areas of the $\kappa$ cutout that are well beyond
the virial radius of the cluster do not contain significant
information about the halo density profile\footnote{Large scales do
  contain information about the halo-matter correlation, which in turn
  is related to the halo mass.  However, our focus here is on
  measuring the halo mass directly in the ``one-halo regime''.}.  We
can therefore speed up our analysis pipeline with little reduction in
signal-to-noise by restricting our analysis to the inner parts of the
$\kappa$ cutouts.  To this end, we restrict our fitting to the inner
$10'\times 10'$ region at the center of the full $\kappa$ cutouts.
The size of this reduced cutout can be compared to the halo virial
radius of a $M = 5\times10^{14} M_{\odot}$ halo at the mean redshift
of the cluster sample ($\bar{z} = 0.45$), which is $\sim 5$
arcminutes.  As noted previously, the richness limit imposed in this
analysis corresponds to roughly $3.25\times 10^{14}\,M_{\odot}$, so
restricting the analysis to the inner 10 arcminutes captures the
region within the virial radius for the majority (if not all) of the
clusters in our sample.  To reiterate: we use large $128'\times128'$
cutouts of the CMB temperature maps to estimate $\kappa$, but we only
use the inner $10'\times 10'$ of the resulting $\kappa$ map for
cluster mass estimation.

Even if the true $\kappa$ in the cutout is zero, the application of
the quadratic estimator to the cutout is still expected to return a
non-zero estimate of $\kappa$ because of the apodization window that
is applied.  To estimate the true $\kappa$, then, we must subtract an
estimate of the mean $\kappa$ in the absence of any CMB lensing,
i.e. the {\it mean field}.  We form an estimate of the mean field for
each observation field by performing the $\kappa$ estimation process
around random locations within the field.  The number of random points
is approximately forty times the number of clusters in each field, and
we confirm that the scatter in the mean field estimate for each field
is negligible compared to the noise in the $\kappa$ estimate around
the clusters.


Because the signal-to-noise for the $\kappa$ measurements around an
individual cluster is much less than one, we form a stack of the
$\kappa$ cutouts.  To maximize the signal-to-noise of the stack, we
use inverse variance weighting when stacking.  The stacked $\kappa$
measurement, $\vec{\kappa}_s$, is then
\begin{eqnarray}
\vec{\kappa}_s = \frac{\sum_i w_i \vec{\kappa_i}}{\sum_i w_i},
\end{eqnarray}
where the sum runs over all cutouts and $w_i = 1/\sigma_i^2$ is the
inverse variance weight.  The estimate of the variance, $\sigma_i^2$,
used for weighting is the same for each field and is calculated by
taking the variance across all $\kappa$ cutouts in that field.  Note
that we do not attempt to perfectly align cluster centers when
performing the stacking; instead, we keep track of the full coordinate
information for each cutout, and take this into account when
constructing the model for the stacked $\kappa$ cutout.  The vector
notation for $\vec{\kappa}_s$ indicates that the measurements are a
function of pixel location across the cutout (defined relative to the
cluster center).

Because the SPT transfer function is anisotropic, the $\kappa$ cutouts
necessarily have anisotropic noise.  We therefore fit the stacked 2D
$\vec{\kappa}_s$ cutout in our analysis rather than the azimuthally
averaged profile of this cutout, as will be described in more detail
below.  To estimate the covariance of $\vec{\kappa}_s$ we use
jackknifing.

\section{Fitting the $\kappa$ measurements}
\label{sec:modelfit}

\subsection{Model}
\label{sec:model}

We fit the 2D stacked $\vec{\kappa}_s$ measurement to constrain the
relation between $M_{200m}$ and $\lambda$ for the clusters in our
sample.  Each cluster is modeled as the sum of a ``1-halo'' term resulting
from the mass of the cluster itself, and a ``2-halo'' term resulting
from correlated structure along the line of sight.

We model the 1-halo term of each cluster using the Navarro,
Frenk and White (NFW) \citep{Navarro:1996} density profile:
\begin{eqnarray}
\label{eq:nfw}
\rho(x) = \frac{200\rho_m(z)}{3} \left[ \frac{c^3(1+c)}{(1+c)\ln(1+c) - c} \right]\frac{1}{x(1+x)^2},
\end{eqnarray}
where $x = rc/R_{200m}$, $c$ is the concentration parameter and $z$ is
the redshift of the cluster.  We set $c$ using the mass-concentration
relation from \citet{Diemer:2015}, but find that our results are
essentially unchanged if the mass-concentration relation from
\citet{Duffy:2008} is used instead.  The projected density along
the line of sight is then
\begin{eqnarray}
\Sigma_{1h}(R) = \int_{-\infty}^{\infty} dh \, \rho\left(r = \sqrt{R^2 + h^2} \right).
\end{eqnarray}
Analytic formulae for $\Sigma_{1h}$ corresponding to the density profile of
Eq.~\ref{eq:nfw} can be found in e.g. \citet{Bartelmann:1996}.

We model the 2-halo term following \citet{Oguri:2011}.  The projected
density profile due to correlated structure along the line of sight is
written as
\begin{multline}
  \Sigma_{2h}\left(\theta = \frac{R}{D_A(z)} \right) = \int \frac{\ell d\ell}{2\pi }\\
  J_0(\ell \theta) \frac{\rho_m(z)b(M)}{(1+z)^3 D_A^2(z)} P_{m}\left(k = \frac{\ell}{(1+z)D_A(z)}, z\right) ,
\end{multline}
where $D_A(z)$ is the angular diameter distance to the cluster, $J_0$
is the zero-th order Bessel function of the first kind, $\rho_m(z)$ is
the mean matter density of the Universe, $P_m(k,z)$ is the linear
matter power spectrum, and $b(M)$ is the clustering bias of halos with
mass $M$.  We model the halo bias using the fitting formulae from
\citet{Tinker:2010}.  The total projected density along the line of
sight is then $\Sigma(R) = \Sigma_{1h}(R) + \Sigma_{2h}(R)$.

The lensing convergence, $\kappa$, is related to the projected density
along the line of sight via $\kappa = \Sigma/\Sigma_c$,
where $\Sigma_c$ is the critical surface density,
\begin{eqnarray}
\Sigma_c = \frac{c^2}{4\pi G} \frac{D_S}{D_L D_{LS}},
\end{eqnarray}
and $D_S$, $D_L$, and $D_{LS}$ are the angular diameter distances to
the source (i.e. the last scattering surface), the lens (i.e. the
cluster), and between the lens and source.  

Because our analysis uses cluster centers determined by redMaPPer, we
must also account for differences between the redMaPPer-identified
center and the true halo center, i.e. miscentering.  We follow an
approach to accounting for miscentering similar to that of
\citetalias{Melchior:2017}.  A cluster that is miscentered by $R_{\rm
  mis}$ will result in a projected density profile given by
\begin{eqnarray}
\label{eq:miscentering}
&&\Sigma_{\rm mis}(R|R_{\rm mis}) = \nonumber \\
&& \int_0^{2 \pi} \frac{d\theta}{2\pi}
\Sigma \left(\sqrt{R^2 + R_{\rm mis}^2 + 2 RR_{\rm mis} \cos \theta} \right)
\end{eqnarray}
where $\Sigma(R)$ is the projected density profile without
miscentering \citep[e.g.][]{Yang:2006}.  

By comparing redMaPPer centers identified in DES SV data measurements
in X-ray and SZ, \citet{Rykoff:2016} constrained the fraction of
miscentered clusters in DES SV data to be $f_{\rm mis} = 0.22 \pm
0.11$.  \citet{Rykoff:2016} modeled the assumed cluster center as
being a draw from a two-dimensional Gaussian with variance
$\sigma_R^2$ centered on the true cluster center.  In this model,
$R_{\rm mis}$ follows a Rayleigh distribution which peaks at
$\sigma_R$.  \citet{Rykoff:2016} further assumed that $\sigma_R$ was
proportional to the \redmap{} defined cluster radius, $R_{\lambda} =
(\lambda/100)^{0.2} h^{-1} {\rm Mpc}$, and constrained $\sigma_R =
c_{\rm mis} R_{\lambda}$ with $\ln c_{\rm mis} = -1.13\pm 0.22$.

In this analysis, we simply assume that a fraction $f_{\rm mis}$ of
the clusters are miscentered by a distance $\sigma_R = c_{\rm mis}
R_{\lambda}$.  The miscentered $\kappa$ profile can then be written as
\begin{eqnarray}
\label{eq:kappa_miscentered}
\kappa(R ; M_{200m},z) = \frac{(1-f_{\rm mis})\Sigma_0(R) + f_{\rm mis} \Sigma_{\rm
    mis}(R)}{\Sigma_c},
\end{eqnarray}
where $\Sigma_0$ is the projected NFW density profile without
miscentering.  In our fiducial analysis, we take the approach of
fixing $f_{\rm mis} = 0.22$ and $\ln c_{\rm mis} = -1.13$, i.e. the
best-fit values from \citetalias{Melchior:2017}.  We will quantify the
uncertainty on our mass-richness constraints that is associated with
the miscentering model in \S\ref{sec:sys_miscentering}.  The final
$\kappa$ model for a cutout is obtained by convolving the miscentered
$\kappa$ map with the filter described in \S\ref{sec:filtered_kappa}.

Following several previous studies constraining the mass-richness
relation of redMaPPer clusters (e.g. \citealt{Simet:2017},
\citealt{Baxter:2016}, \citetalias{Melchior:2017}), we adopt a power
law relation for the expectation value of the mass, $M$, at fixed richness
and redshift:
\begin{eqnarray}
\label{eq:mr}
\langle M | \lambda_i , z_i ; \vec{p}_{\rm m.r.} \rangle = A \left(
\frac{\lambda}{\lambda_0}\right)^{\alpha} \left(
\frac{1+z}{1+z_0}\right)^{\beta},
\end{eqnarray}
where we fix the pivot points at $\lambda_0= 30$ and $z_0 = 0.5$ to
match \citetalias{Melchior:2017}.  The parameters of the mass-richness
relation are the amplitude, $A$, the richness scaling, $\alpha$, and
the redshift scaling, $\beta$; we denote the vector of these
parameters with $\vec{p}_{\rm m.r.}$.

Given $\vec{p}_{\rm m.r.}$, our model for the stacked $\kappa$
profile, $\vec{\kappa}_s^m$, is then the weighted average across all
clusters of $\kappa$ evaluated at the expectation value of the mass
for each cluster:
\begin{eqnarray}
\label{eq:kappa_model}
\vec{\kappa}_s^m(\vec{p}_{\rm m.r.} ) = \frac{\sum_i w_i \vec{\kappa}(
  \langle M | \lambda_i , z_i; \vec{p}_{\rm m.r.} \rangle,
  z_i)}{\sum_i w_i},
\end{eqnarray}
where the $w_i$ are the inverse variance weights introduced in
\S\ref{sec:filtered_kappa} and $\vec{\kappa}$ is given by the
mis-centered model of Eq.~\ref{eq:kappa_miscentered} after application
of the filter described in \S\ref{sec:filtered_kappa}.  Below, we will
fit the stacked cluster profile $\vec{\kappa}_s$ with the three
parameter model defined in this section.

\subsection{Likelihood analysis}
\label{sec:likelihood_analysis}

We now describe the process of fitting the stacked $\kappa$ cutout to
obtain constraints on the mass-richness relation of the \redmap{}
clusters.  We begin by assuming a Gaussian likelihood for the data
vector, $\vec{\kappa}_s$, given the model $\vec{\kappa}^m_s$ from
Eq.~\ref{eq:kappa_model}:
\begin{eqnarray}
\ln \mathcal{L}(\vec{\kappa}_s | \vec{p}_{\rm m.r.}; \{\lambda_i, z_i \})  =  -\frac{1}{2} \left[ \vec{\kappa}_s - \vec{\kappa}_s^m \right]^T   \mathbf{C}^{-1} \left[ \vec{\kappa}_s - \vec{\kappa}_s^m\right], \nonumber \\
\label{eq:likelihood} 
\end{eqnarray}
where $\mathbf{C}$ is the covariance matrix of the $\vec{\kappa}_s$
measurement estimated using jackknife resampling.

The posterior on the mass-richness parameters can then be written as
\begin{eqnarray}
\label{eq:posterior_mr}
\mathcal{P}(\vec{p}_{\rm m.r.} | \vec{\kappa}_s ; \{ z_i \}) =
 \mathcal{L}(\vec{\kappa}_s | \vec{p}_{\rm m.r.}; \{\lambda_i, z_i \}) 
\mathcal{P}(\vec{p}_{\rm m.r.}), \nonumber \\
\end{eqnarray}
where the last term, $\mathcal{P}(\vec{p}_{\rm m.r.})$, represents the
priors on the mass-richness parameters.  Given the signal-to-noise of
our measurements, we do not expect to be able to robustly constrain
many parameters.  We therefore focus on constraining the amplitude of
the mass-richness relation, $A$.  We impose a flat prior on $A$
between $[10^{12},10^{16}]$, but find that our results are quite
insensitive to the form of this prior, indicating that our constraints
are dominated by information in the likelihood.  We impose a Gaussian
prior on $\alpha$ motivated by the results of
\citetalias{Melchior:2017} and \citet{Simet:2017} (hereafter
\citetalias{Simet:2017}).  These two analyses found $\alpha = 1.12 \pm
0.20$ and $1.33\pm0.1$, respectively.  For the central value of our
prior on $\alpha$, we simply take the average of these two values.
For the width of the prior, we assume $\sigma_{\alpha} = 0.3$.  This
value is meant to reflect the statistical uncertainty in $\alpha$
along with any uncertainty owing to differences in the definition of
richness between \citetalias{Melchior:2017}, \citetalias{Simet:2017},
and Y1 DES data.  As we will show below, our constraint on $A$ is not
very degenerate with $\alpha$, so having a relatively loose prior on
$\alpha$ is acceptable.  To simplify the analysis, we fix $\beta =
0.18$ (equivalent to assuming a $\delta$-function prior on this
parameter), corresponding to the best-fit value from
\citetalias{Melchior:2017}.  We find that our analysis is almost
entirely insensitive to $\beta$, so fixing this parameter has
essentially no impact on our results. These priors are summarized in
Table~\ref{tab:mr_parameters}.  The resultant parameter space is
two-dimensional ($A$ and $\alpha$) and can be explored using a simple
grid sampler.

Eq.~\ref{eq:posterior_mr} ignores several potential sources of
uncertainty in the stacked model, $\vec{\kappa}_s^m$.  First, we have
ignored scatter in the mass-richness relation.  This scatter is
expected to be described by a log-normal probability distribution
function, with scatter at fixed $\lambda$ given by $\sigma_{\ln M |
  \lambda} \sim 0.25$ \citep[e.g.][]{Rozo:2014}.  At low richness, one
expects increased scatter due to the Poisson uncertainty in the number
of galaxies.  However, even accounting for the additional Poisson
scatter (following the prescription described in \citetalias{Simet:2017})
we find that the uncertainty on $\vec{\kappa}_s^m$ is much less than
the uncertainty on the measurement vector, $\vec{\kappa}_s$.
Consequently, without introducing any measurable bias in our results,
we can ignore scatter in the mass-richness relation.

Additionally, Eq.~\ref{eq:posterior_mr} ignores uncertainty in the
redshift and richness estimates for the clusters.  The typical
redshift uncertainty for DES \redmap{} clusters is $\sigma_z/(1+z)
\lesssim 0.01$ \citep{Rykoff:2016}.  For this level of redshift
uncertainty, the resultant uncertainty in $\vec{\kappa}_s^m$ is much
less than the uncertainty in the $\kappa$ measurements, so it is safe
to ignore redshift uncertainty in this analysis.  A similar argument
holds for the uncertainty on the richness estimates.

Finally, note that Eq.~\ref{eq:posterior_mr} ignores correlated
scatter between the richness and lensing mass.  Such correlated
scatter is expected because clusters that are elongated along the line
of sight will have enhanced richness and lensing masses.  However, the
impact of such correlated scatter is expected to be at the few percent
level \citep{Melchior:2017}, well below the statistical uncertainties
obtained here.  For simplicity we therefore ignore this effect.

\section{Pipeline Validation}
\label{sec:simulations}

We use simulated SPT observations to test the $\kappa$ estimation
pipeline.  We generate simulated SPT maps by adding together Gaussian
realizations of the CMB, foreground and noise.  The CMB realizations
are generated from a power spectrum computed at our fiducial
cosmological model using CAMB\footnote{\texttt{http://camb.info}}
\citep{Lewis:1999}.

We lens the simulated CMB maps with mock clusters described by the
NFW profile of Eq.~\ref{eq:nfw}.  To perform the lensing operation, we
compute the deflection angles for these clusters using the formulae of
\citet{Bartelmann:1996}.  The unlensed CMB is then remapped to the
lensed CMB with cubic spline interpolation.  The main purpose of the
simulations is to confirm that our lensing pipeline recovers an
unbiased estimate of $\kappa$ to within our noise levels.  We
therefore use clusters of fixed mass and redshift when generating the
simulations.  We use $100$ mock galaxy clusters that are roughly
equally spaced across the field.  Each simulated cluster has $M =
2.5\times10^{14}\,M_{\odot}$, $z = 0.4$ and $c = 4.8$.  This mass is
somewhat higher than the mean mass predicted for the \redmap{} sample
based on the mass-richness relation of \citetalias{Melchior:2017},
which is $M \sim 2.05\times 10^{14}\,M_{\odot}$.

Foreground emission is generated as Gaussian random realizations of
the foreground models described in \S\ref{sec:noise_estimation} and
added to the lensed CMB maps.  We assume that foreground emission is
unlensed by the galaxy clusters when generating the simulated maps
because this matches the assumption of our fiducial analysis; we
consider the effects of this assumption on the analysis of the real
data in \S\ref{sec:foreground_lensing_systematic}.  Note that for the
purposes of pipeline validation, we do not include tSZ signal in the
simulated data; we estimate the effects of tSZ contamination in
\S\ref{sec:tsz}.

The simulated sky maps are then convolved with the beam and
field-dependent transfer functions from \citetalias{George:2015}.
Finally, field-dependent noise realizations are added to the simulated
maps.  To generate the field-dependent noise realizations, we generate
Gaussian random realizations of the estimated noise power spectrum
from the weight-scaled difference maps described in
\S\ref{sec:noise_estimation}.  The noise realizations are then scaled
by the inverse square root of the weight to account for weight
variations across the field.

To build statistics, we generate $200$ simulated skies using the
methods described above.  Each simulated sky map is lensed with 100
clusters, bringing the total number of simulated clusters to 20,000.
Each simulated sky has a different random realization of the CMB,
foregrounds and noise.  The simulated skies are passed through the
same pipeline that is applied to the data to extract $\kappa$ cutouts
around each of the simulated clusters.  The $\kappa$ cutouts are then
fit to determine constraints on the cluster mass, $M_{200m}$.

\begin{figure}
\includegraphics[width = \columnwidth]{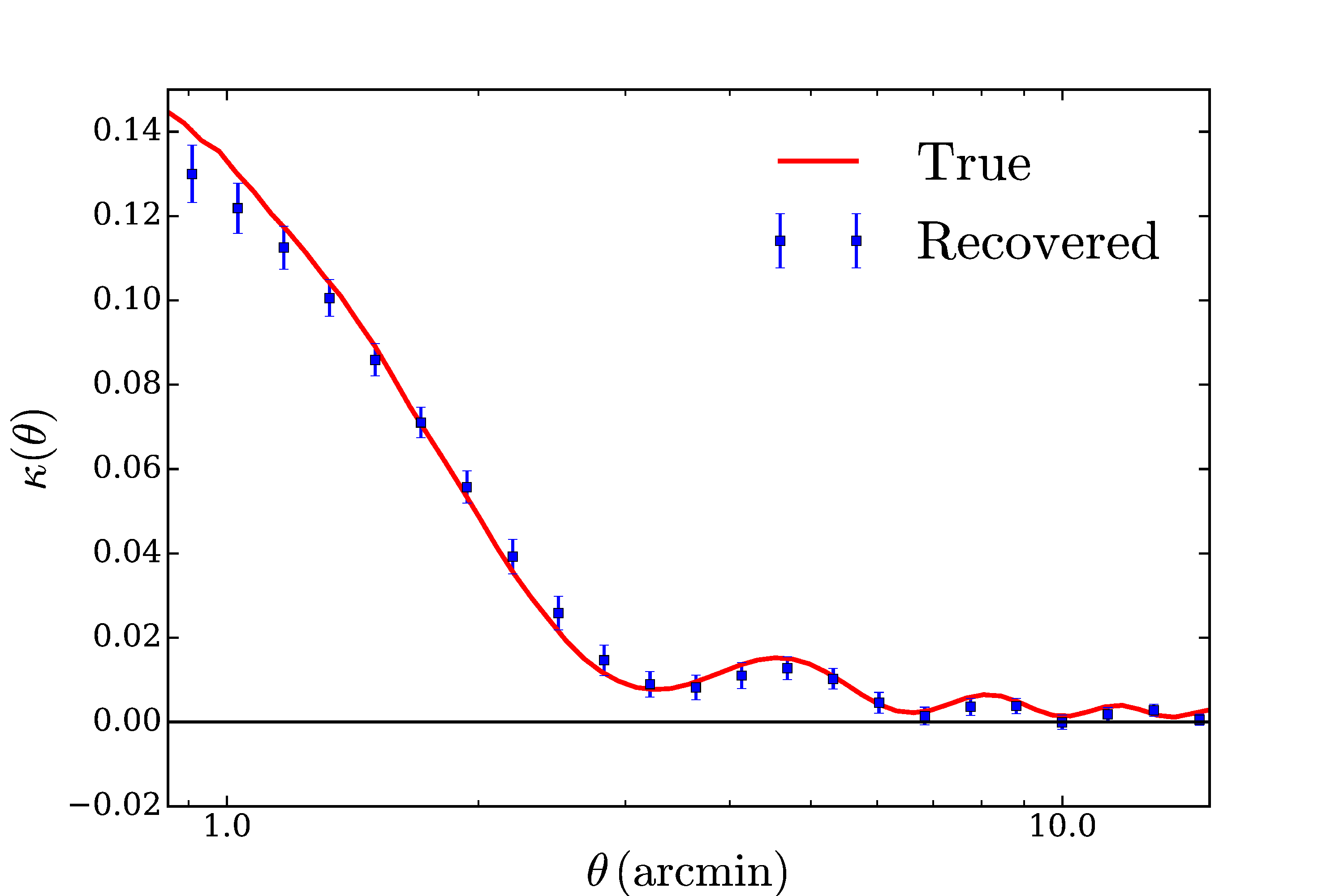}
\caption{Azimuthally averaged $\kappa$ profile recovered from analysis
  of simulated data.  Red solid curve shows the true $\kappa$ profile
  around mock clusters with $M = 2.5\times10^{14}\,M_{\odot}$ after
  the application of the filtering described in
  \S\ref{sec:filtered_kappa}.  Blue data points with errorbars show
  recovered $\kappa$ in the presence of realistic noise, foreground
  emission, beam, and transfer function using 20,000 clusters.  The
  measurement pipeline recovers an unbiased estimate of the true
  $\kappa$ profile to within the uncertainties.}
\label{fig:pipeline_test}
\end{figure}

Fig.~\ref{fig:pipeline_test} shows the true, azimuthally averaged
$\kappa$ profile (red) in the simulations compared to the $\kappa$
profile recovered using our analysis pipeline (points with error
bars).  The oscillatory behavior of $\kappa$ comes from application of
the filter described in \S\ref{sec:filtered_kappa} to the $\kappa$
cutouts.  The pipeline recovers an unbiased estimate of $\kappa$ to
within the uncertainties of the mock data.  Note that we have used
roughly five times as many simulated clusters as real clusters for
this test to increase our sensitivity to any possible biases in the
$\kappa$ estimation.

\begin{figure}
\includegraphics[width = \columnwidth]{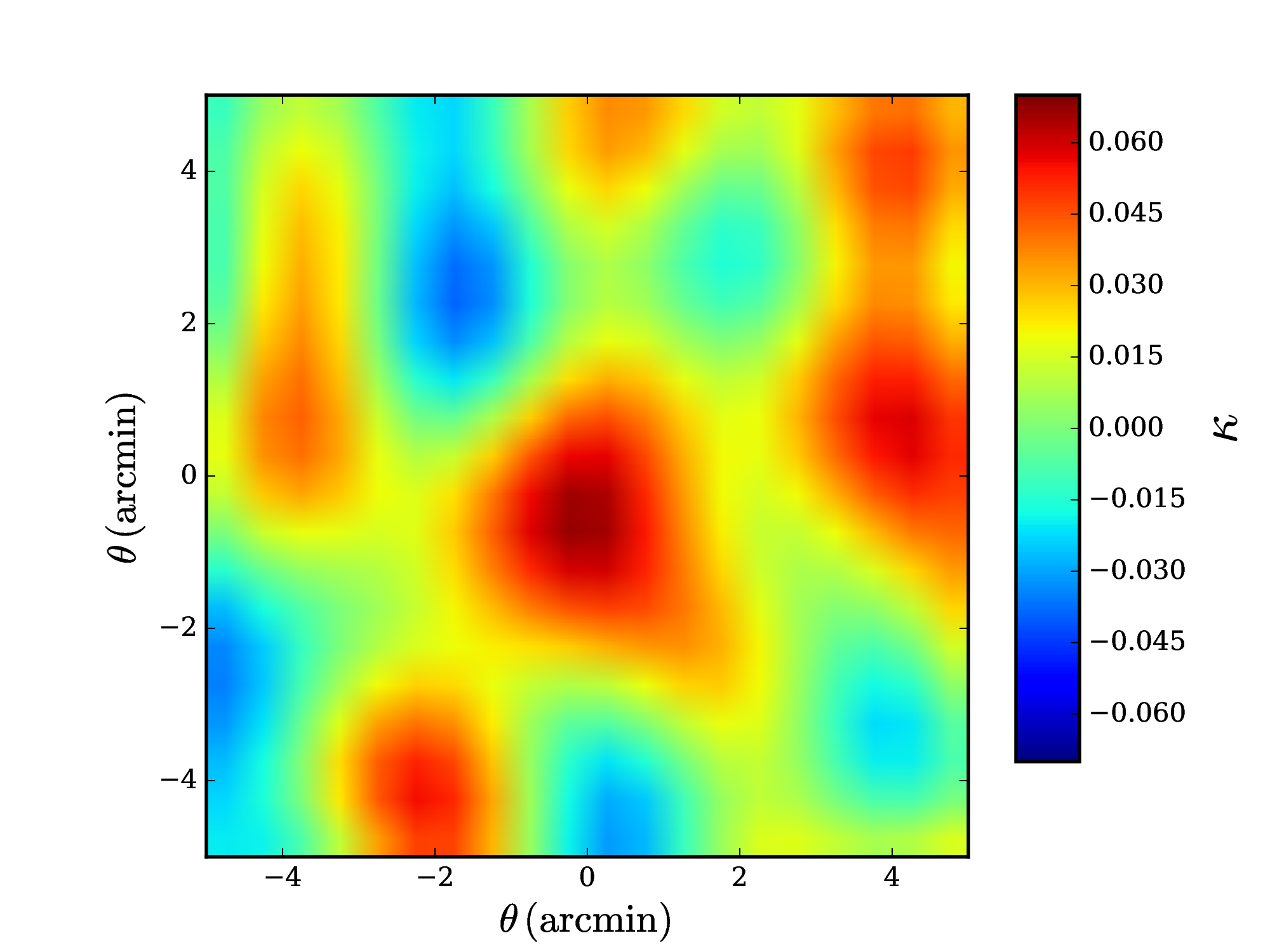}
\caption{The stacked, weighted 2D $\kappa$ profile recovered from the
  analysis of CMB temperature data around 3697 \redmap{} clusters.
  Each pixel in the cutout is 0.5 arcminutes on a side.}
\label{fig:2d_profile_data}
\end{figure}

\begin{figure*}
\includegraphics[width = 1.5\columnwidth]{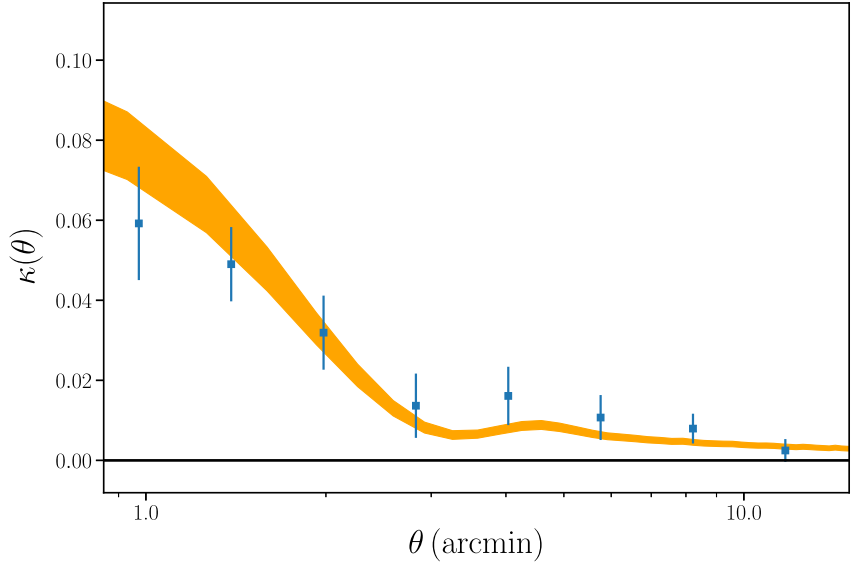}
\caption{The azimuthally averaged $\kappa$ profile recovered from
  measurement of CMB lensing around 3697 \redmap{} clusters detected
  in DES Y1 data (blue points with errorbars).  The errorbars shown
  are the diagonal elements of the full covariance determined from a
  jackknife resampling of the cluster sample; there is significant
  covariance between adjacent data points.  The orange region indicates
  the allowed range of model predictions (68\% confidence region)
  given the results of the fit for parameters of the cluster
  mass-richness relation.}
\label{fig:1d_profile_data}
\end{figure*}

\section{Results}
\label{sec:results}

\subsection{$\kappa$ measurements around redMaPPer clusters}

Fig.~\ref{fig:2d_profile_data} shows the weighted average of the
$\kappa$ cutouts around \redmap{} clusters, restricted to the region
used in fitting.  Pixels in the $\kappa$ cutouts are 0.5 arcminutes on
a side and the fitted region is 20 pixels by 20 pixels.  Because the
centers of the \redmap{} clusters do not lie at exactly the same
position in each map pixel, Fig.~\ref{fig:2d_profile_data}
incorporates some smearing due to pixelization effects.  However, when
fitting for the parameters of the mass-richness relation we use the
full coordinate information for each pixel relative to the true
cluster centers on a cluster-by-cluster basis.\footnote{A small offset
  in the peak of the recovered convergence map relative to the origin
  may be observed in Fig.~\ref{fig:2d_profile_data}, which we
  attribute to noise.  A similar effect is seen when applying the
  quadratic estimator to the simulations described in
  \S\ref{sec:simulations}; such an offset was also observed in
  \citet{Madhavacheril:2015}.  Averaged over many realizations,
  though, we correctly recover the input convergence maps in the
  simulations.}

Fig.~\ref{fig:1d_profile_data} shows the azimuthally averaged
one-dimensional $\kappa$ profile extracted from the analysis of
\redmap{} clusters.  To determine the error bars shown in this plot we
use a jackknife resampling approach, where the jackknife subsamples
are determined by dividing survey area into 100 regions of
approximately equal area.  The data exhibit a strong preference for
increasing $\kappa$ towards the center of the cluster, as expected.
We note, though, that adjacent measurements in
Fig.~\ref{fig:1d_profile_data} are highly correlated.  While we show
the 1D $\kappa$ profile here for the purposes of visualization, our
analysis to extract mass constraints on the clusters uses the full 2D
$\kappa$ information, rather than the 1D azimuthally averaged profile.

\subsection{Fit results}
\label{sec:fit_results}

\begin{figure}
\includegraphics[width = \columnwidth]{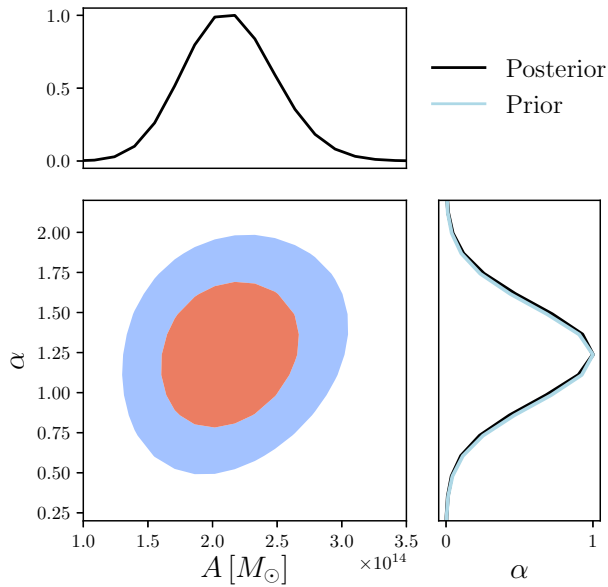}
\caption[Caption for multi posterior figure]{Constraints on the
  amplitude, $A$, and richness scaling, $\alpha$, of the \redmap{}
  mass-richness relation obtained from fits to CMB lensing
  measurements.  Contours represent $1\sigma$ and $2\sigma$ levels.
  The blue curve in the panel at right shows the prior on $\alpha$,
  which dominates our constraint.  Numerical results are summarized in
  Table~\ref{tab:mr_parameters}.}
\label{fig:mr_multid}
\end{figure}

The 2D and marginalized posteriors on the mass-richness parameters $A$
and $\alpha$ recovered from our analysis are shown in
Fig.~\ref{fig:mr_multid}; numerical results are given in
Table~\ref{tab:mr_parameters}.  We find $A/M_{\odot} = (2.14 \pm
0.35)\times 10^{14}$, a roughly 17\% constraint on the amplitude of the
mass-richness relation.  The posterior on $\alpha$ is entirely
dominated by the prior (shown as the blue curve).  This is not
surprising given the fairly low signal-to-noise of our measurement and
the fact that the richness range of the sample is restricted to $20 <
\lambda < 40$.  There appears to be minimal degeneracy between $A$ and
$\alpha$ over the range of $\alpha$ allowed by the constraints of
\citetalias{Melchior:2017} and \citetalias{Simet:2017}.  If we fix
$\alpha$ to the best-fit value from \citetalias{Melchior:2017}, we
find that the best-fit $A$ changes by only 1\% and the error on $A$
decreases by only 2\%.  Because our prior on $\alpha$ is quite wide,
our results are therefore very robust to assumptions about $\alpha$.

We find $\Delta \ln \mathcal{L} = 33.3$ between the peak of the
likelihood and $A = 0$; a likelihood ratio test therefore allows us to
reject the null hypothesis that $A = 0$ at $8.1\sigma$ significance.
The significance of this detection can be compared to the $3.1\sigma$
detection obtained in B15 from 513 clusters.  Those clusters were
approximately twice as massive as the ones considered here, so we
would expect that despite the increase in sample size our current
measurement would yield a $\sim4.0\sigma$ detection.  The higher
detection significance reported here is the result of the switch from
the tSZ-free maps used in B15 to the lower-noise 150~GHz SPT maps that
we are able to use with the quadratic estimator in the current
analysis.

\begin{table}
\centering
\caption{Constraints on the mass-richness parameters.  Top-hat
  distributions are indicated by brackets, while Gaussian
  distributions are written in the form $a \pm b$.}
\label{tab:mr_parameters}
\begin{tabular}{|c|c|c|c|}
\hline
Parameter & Prior & Posterior  \\
$A/M_{\odot}$ & $[10^{12}, 10^{16} ]$ & $(2.14\pm0.35)\times 10^{14} M_{\odot}$ \\ 
$\alpha$ & $1.23 \pm 0.30$ & $1.25 \pm 0.30$ \\
$\beta$ & $0.18$ & --- \\
\hline
\end{tabular}
\end{table}

\subsection{Systematic errors}
\label{sec:systematics}

\subsubsection{tSZ}
\label{sec:tsz}

The tSZ effect is caused by inverse Compton scattering of CMB photons
with energetic electrons.  The effect is especially pronounced in the
direction of massive galaxy clusters, as these objects are reservoirs
of hot gas.  At 150 GHz --- the frequency of observation for the CMB
maps used in this analysis --- the tSZ effect leads to a decrement in
the observed CMB temperature near the cluster \citep[for a review
  see][]{Birkinshaw:1999}.  We do not attempt to model the tSZ in this
analysis; consequently, its presence acts as a potential source of
bias to our measurement of $\kappa$.  Somewhat worryingly, the
magnitude of the tSZ decrement for a massive cluster can be $\sim
100\,\mu{\rm K}$, significantly larger than the magnitude of the CMB
cluster lensing-induced distortion, which has amplitudes $\lesssim
10\,\mu{\rm K}$.  However, the situation is not so bad as it might
first appear: unlike the CMB cluster lensing signal, the tSZ effect is
not correlated with the CMB gradient behind the cluster.  Because the
quadratic estimator used to measure the lensing distortion is
effectively picking out correlations between small scale CMB
distortions and the larger scale gradient field, it is expected to be
fairly robust to tSZ contamination.  Furthermore, the low-pass
filtering imposed on the CMB maps to estimate the gradient field
(\S\ref{sec:quad_estimator}) effectively filters out the small-scale
tSZ decrements, reducing their contamination of the $\kappa$
estimator.  Finally, as noted in \S\ref{sec:cluster_catalog}, we
restrict our analysis to clusters with $\lambda < 40$ to reduce the
amplitude of tSZ contamination.

To constrain the level of systematic error introduced to our $\kappa$
estimates by the tSZ we rely on simulations.  We introduce mock tSZ
signals into the simulations described in \S\ref{sec:simulations} and
re-analyze these simulations to determine how much the $\kappa$
estimation is biased.  The mock tSZ signals used for this purpose are
taken from the hydrodynamical simulations of \citet{LeBrun:2014}.
These simulations represent an extension of the OverWhelmingly Large
Simulations project \citep{Schaye:2010}, and are designed with
applications to cluster cosmology in mind.  To this end, the
simulations are large volume (400 $h^{-1}$ Mpc on a side) and include
computation of maps of the Compton $y$ parameter.  We use their AGN
8.0 model in this analysis.

To introduce tSZ into our simulations, we extract tSZ cutouts
measuring 256 arcminutes on a side from the \citet{LeBrun:2014}
Compton $y$ maps at the locations of massive halos.  We restrict the
cutouts to those halos with $1.75\times 10^{14} M_{\odot} < M <
3.25\times 10^{14} M_{\odot}$ and $0.24 < z < 0.56$, as these should
be well matched to the real clusters.\footnote{Halo masses in
  $M_{500c}$ are converted to $M_{200m}$ assuming that the mass
  follows an NFW profile and using the mass-concentration relation from
  \citet{Duffy:2008}.}  A cluster with mass $M_{200m} \sim 3.25\times
10^{14} M_{\odot}$ corresponds roughly to a richness of $\lambda \sim
40$ assuming the mass-richness relation of \citetalias{Melchior:2017}.
While some clusters with $\lambda \sim 40$ in our sample may have
masses larger than $3.25\times 10^{14} M_{\odot}$ owing to scatter in
the mass-richness relation, we expect the constraint to be dominated
by clusters below this mass.  Comptonization maps are converted to
temperature units assuming an observation frequency of 150 GHz.  This
process yields 135 simulated tSZ cutouts.

The simulated tSZ cutouts are added to simulated temperature maps at
the locations of the mock clusters.  We perform the $\kappa$
estimation process on 2700 mock clusters with added mock tSZ signal.
For each cluster, the tSZ signal is rotated by a random angle.  With
the mock tSZ profiles added to the simulations, the application of the
beam, transfer function and noise then proceeds as before and the
cutouts are processed through the $\kappa$ estimation pipeline.  The
results of this analysis are then compared to a set of simulations
that are identical in every way (i.e. the same realizations of CMB,
noise, and foregrounds) except they do not have the added tSZ
component.

Analyzing the simulated cutouts with the mock tSZ signal reveals that
the recovered mass in the presence of tSZ is biased low relative to
the mass in the absence of tSZ by a few percent.  This level of bias
is significantly smaller than the statistical errors associated with
our $\kappa$ estimates.

There is some systematic uncertainty in the simulated tSZ profiles
resulting from differences between the \citet{LeBrun:2014} simulations
and real galaxy clusters.  \citet{LeBrun:2014} showed that their
simulations were able to accurately reproduce the relation between
integrated Compton parameter and mass for a set of nearby galaxy
clusters observed by \citet{Planck:2011} and \citet{Planck:2013}.
However, there is significant scatter between the different `sub-grid'
physics models considered by \citet{LeBrun:2014}.  To account for
this, we repeat the process of introducing the simulated tSZ profiles
into the mock data after increasing the amplitude of the simulated tSZ
profiles by 30\%.  With this increased tSZ amplitude, we find that the
bias in the recovered mass estimate increases to roughly 11\%.  Again,
this level of bias is smaller than our statistical error bar, but is
certainly non-negligible.  We emphasize that this bias is one-sided:
it acts to reduce the inferred amplitude of the recovered $\kappa$
estimate.

We have also repeated the above analysis using a more aggressive
gradient filter scale of $l_G = 2000$.  This choice yields a higher
signal to noise reconstruction of the cluster profile in the absence
of tSZ.  However, in the presence of tSZ (with amplitude fixed to that
of the \citealt{LeBrun:2014} simulations), choosing $l_G = 2000$ results
in recovered mass estimates that are biased low by as much as 30\%.
Our fiducial choice of $l_G$ therefore has the effect of significantly
reducing the bias due to tSZ at the cost of increasing our error bars
somewhat.

As a further test of tSZ contamination, we also consider the
effect of varying the maximum richness threshold imposed in our
analysis of the \redmap{} clusters.  Because the amplitude of the tSZ
signal for a cluster of mass $M$ scales roughly as $M^{5/3}$ while the
lensing signal is roughly proportional to $M$, we expect high-richness
clusters to be more impacted by tSZ bias.  Indeed, we find that very
high richness clusters ($\lambda > 100$) tend to exhibit a preference
for low masses.  One cluster with richness $\lambda \sim 180$ in
particular exhibits a fairly significant preference for negative mass.
As a result, if we include all the clusters in the analysis, the
preference for $A > 0$ actually decreases slightly.  However, when
varying the richness threshold between $40 < \lambda \lesssim 50$,
the preference for $A > 0$ tends to increase with increasing richness
threshold.  This suggests that tSZ contamination is fairly minimal for
clusters in this richness range.

Another way to constrain the presence of tSZ bias in our analysis is
to look at the posterior on $\alpha$.  Because high richness clusters
are expected to have their $\kappa$ estimates biased low by the
presence of tSZ, if such bias were significant in our measurement, we
would expect the recovered posterior on $\alpha$ to prefer low values.
Fig.~\ref{fig:mr_multid} shows both the prior and posterior on
$\alpha$.  The posterior on $\alpha$ is entirely consistent with the
prior, suggesting that $\alpha$ is not being driven low by tSZ bias.

In addition to the tSZ effect, galaxy clusters are also expected to
distort the CMB via the kinematic SZ effect (kSZ), caused by inverse
Compton scattering of CMB photons with electrons that have large bulk
velocities relative to the CMB frame.  The kSZ is expected to be
significantly smaller than the tSZ in most cases, and like the tSZ it
is uncorrelated with the CMB gradient behind the cluster.
Furthermore, since the sign of the kSZ signal depends on the direction
of the cluster peculiar velocity relative to the line-of-sight
direction, its effects should be suppressed in an average across many
clusters.  Consequently, the impact of kSZ on our analysis is expected
to be negligible.

\subsubsection{Foreground lensing}
\label{sec:foreground_lensing_systematic}

Our fiducial analysis assumes that all foreground emission is unlensed
by the galaxy clusters.  However, as described in
\S\ref{sec:noise_estimation}, some foreground emission may be
sourced from behind the cluster and will therefore be gravitationally
lensed by the cluster.  To determine the bias introduced into our
analysis by the assumption of unlensed foreground, we repeat the
analysis of the data assuming instead that all foreground emission
originates from the surface of last scattering, and is therefore
maximally lensed.  In that case, the power spectrum of the foreground
emission can be simply added to the CMB power spectrum when computing
the quadratic estimator.  These two extreme assumptions --- no
foreground lensing or maximal foreground lensing --- bracket the
possible levels of foreground lensing, and the difference between the
two resultant $\kappa$ estimates provides an (over)estimate of the
systematic error introduced into our analysis by our foreground
lensing assumption.

When we repeat the analysis of the data with the alternate foreground
lensing assumption, we find that the change to the resultant mass
constraints is less than $1\%$, well below our statistical
uncertainty.

\subsubsection{Miscentering}
\label{sec:sys_miscentering}

Our analysis assumes the best-fit values for the miscentering
parameters from \citet{Rykoff:2016}: $f_{\rm mis} = 0.22$ and $\ln
c_{\rm mis} = -1.13$.  However, the \citet{Rykoff:2016} constraints on
the miscentering parameters carry non-negligible uncertainty.  To
quantify the impact of this uncertainty on our mass-richness
constraints, we repeat our analysis with the miscentering parameters
increased an amount equal to the $1\sigma$ uncertainties from
\citet{Rykoff:2016}: $\sigma(f_{\rm mis}) = 0.11$ and $\sigma(\ln
c_{\rm mis}) = 0.22$.

We find that perturbing $f_{\rm mis}$ and $\ln c_{\rm mis}$ by these
uncertainties results in changes to $A$ of 3\% and 4\%, respectively.
Adding these uncertainties in quadrature, we therefore introduce a
5\% systematic uncertainty to our mass constraints to account for
uncertainty on \redmap{} miscentering.  This level of uncertainty is
roughly $29\%$ of our statistical uncertainty.

It may be surprising that uncertainty on the miscentering
parameters introduces such a large systematic error on our mass
constraints.  Using cluster-galaxy lensing and a similar miscentering
model, \citetalias{Melchior:2017} found that miscentering introduced
less than a 1\% error on the normalization of the mass-richness
relation.  This difference in amplitude can be understood as
resulting from the different angular scales that contribute to the two
constraints.  As seen in Fig.~\ref{fig:1d_profile_data}, the inner
most data points exclude $M = 0$ with high significance, suggesting
that most of our constraining power is coming from small angular
scales where miscentering can have a large impact.  The constraint
from \citetalias{Melchior:2017}, on the other hand, receives a large
contribution from larger scales at which miscentering is unimportant.
Indeed, Fig.~11 of \citetalias{Melchior:2017} shows that most of their
signal comes from $R > 1\,{\rm Mpc}$, where the effects of
miscentering are essentially negligible.

\section{Discussion}
\label{sec:discussion}

\subsection{Cluster mass constraint from CMB cluster lensing}

We have presented a $8.1\sigma$ detection of CMB cluster lensing
around \redmap{} clusters.  Our analysis relied on CMB temperature
maps from the SPT-SZ survey and a sample of optically-selected galaxy
clusters identified in Y1 DES imaging.  By fitting the CMB-$\kappa$
measurements around the \redmap{} clusters, we constrained the
amplitude of the mass richness relation to roughly 17\% statistical
precision.

Our systematics analysis suggests that the dominant systematics
affecting our constraints on the \redmap{} mass-richness relation
are cluster miscentering and the presence of tSZ.  Cluster
miscentering contributes a roughly 5\% systematic error to our mass
constraints.  Our analysis attempts to minimize bias due to the tSZ by
restricting the cluster sample to $\lambda < 40$ and applying a
conservative filter when estimating the CMB temperature gradient
field.  To estimate residual bias due to the presence of tSZ we
analyze simulated data with tSZ profiles taken from the AGN 8.0
simulations of \citet{LeBrun:2014} and also employ a data-only
consistency test.  Both of these tests suggest that there is
negligible bias in our analysis due to the tSZ.  However, after
adjusting the simulations to (conservatively) account for uncertainty
in the mock tSZ profiles, we find that the tSZ-caused bias increases
to 11\%.  Note that this bias acts to {\it reduce} the inferred
amplitude of the mass-richness relation, and so should not be thought
of as a two-sided uncertainty. 

\begin{figure}
\includegraphics[width = \columnwidth]{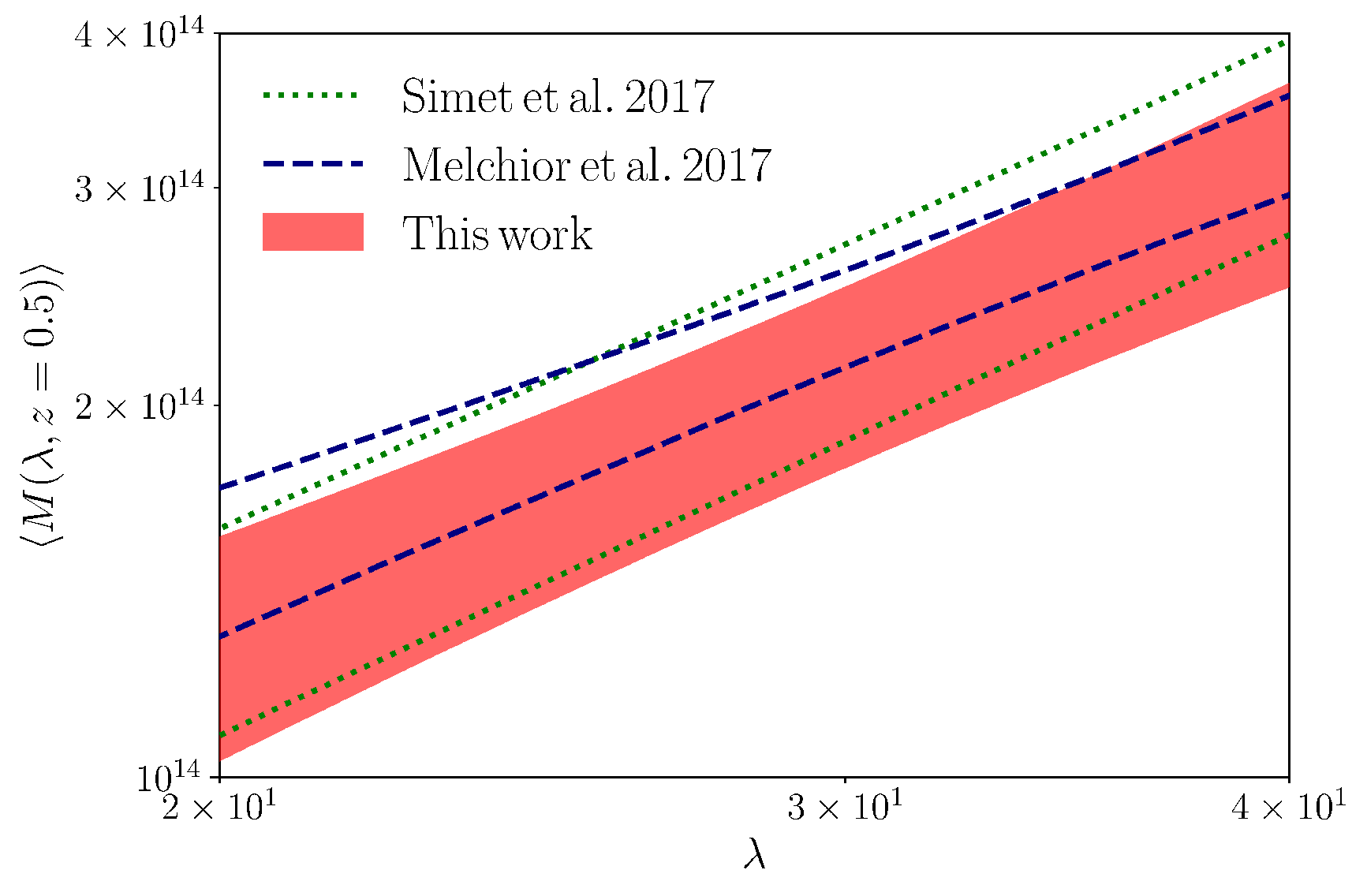}
\caption[Caption for mass-richness comparison figure]{Comparison of
  constraints obtained on the \redmap{} mass-richness relation from
  CMB cluster lensing (this work) and cluster-galaxy lensing
  (\citealt{Melchior:2017} and \citealt{Simet:2017}).  The solid
  band and lines illustrate the $1\sigma$ ranges allowed by the
  different constraints.  Note that unlike the
  \citet{Melchior:2017} and \citet{Simet:2017} analyses,
  this work imposes an informative prior on the slope of the
  mass-richness relation.}
\label{fig:mr_compare}
\end{figure}

\subsection{Comparison to other \redmap{} lensing measurements}
\label{sec:comparison}

The results presented in this work represent the first weak-lensing
mass calibration of the DES Y1 \redmap{} clusters, and therefore a
direct comparison of these results to other weak lensing measurements
with the same cluster sample is not yet possible.  However, given that
the \redmap{} mass-richness relation is expected to be
survey-independent to a good approximation, it is possible to compare
our results to other recent \redmap{} weak lensing mass constraints.

Fig.~\ref{fig:mr_compare} compares the constraint obtained here on the
cluster mass-richness relation to other constraints obtained from
cluster-galaxy lensing.  We evaluate all of the various mass-richness
relations at $z = 0.5$, the pivot redshift of this analysis and
that of \citetalias{Melchior:2017}.  The most direct comparison to
this work is with the weak lensing mass calibration of DES SV clusters
by \citetalias{Melchior:2017}, since that work used the same telescope
and similar modeling assumptions.  Our constraint on the \redmap{}
mass-richness relation is in good agreement with the constraint
from \citetalias{Melchior:2017} over the richness range considered in
this work.  We prefer a slightly lower normalization of the
mass-richness relation, but this difference is not statistically
significant.  Note that our anaylsis uses $\sim 3700$ clusters, while
the \citetalias{Melchior:2017} analysis used only $\sim 1600$
clusters.  Our constraint on the mass-richness relation is also in
good agreement with recent constraints on the mass-richness relation
of \redmap{} clusters in SDSS by \citetalias{Simet:2017}, shown as the
yellow band in Fig.~\ref{fig:mr_compare}.  To translate the
\citetalias{Simet:2017} constraint from the mean redshift of the SDSS
sample to $z = 0.5$, we have assumed the redshift scaling and
uncertainty from the \citetalias{Melchior:2017} analysis, i.e. $\beta
= 0.18 \pm 0.75$.  Were we to instead evaluate the
\citetalias{Simet:2017} mass-richness relation at the mean redshift of
the clusters in that work, the uncertainty on the mass-richness
relation would be significantly reduced.  Finally, we note that unlike
the analyses of \citetalias{Melchior:2017} and
\citetalias{Simet:2017}, we have imposed an informative prior on
$\alpha$ in our analysis (see Table~\ref{tab:mr_parameters}).
However, as shown in Fig.~\ref{fig:mr_multid}, $\alpha$ is not
significantly degenerate with $A$, given the prior used on $\alpha$.

\subsection{Future prospects}

As noted in the introduction, CMB cluster lensing can be used to
provide constraints on systematic errors associated with other weak
lensing cluster mass constraints.  This is possible because the
dominant systematics associated with CMB cluster lensing are expected
to be essentially uncorrelated with those of galaxy lensing
measurements (with miscentering being a notable exception).  The weak
lensing mass calibration of \redmap{} clusters in
\citetalias{Melchior:2017} obtained a systematic uncertainty of roughly
5\%.  Their systematic error budget was dominated by photometric
redshift uncertainty and shear calibration uncertainty, neither of
which affect CMB cluster lensing.  Reaching a 5\% mass calibration
with CMB cluster lensing would require a factor of four improvement in
the statistical uncertainties presented here (ignoring, for a moment, the
contributions of systematic errors).  As we discuss below, such
improvements are certainly possible with future data.

One source of systematic uncertainty that is potentially worrying for
CMB cluster lensing analyses is the presence of tSZ.  Our analysis has
shown that if one considers only clusters with masses below about
$3.3\times10^{14} M_{\odot}$, the bias introduced into the cluster
mass constraints by the tSZ signal --- after adopting conservative
filtering choices --- is at most $\sim 10\%$.  This is an acceptable
level of bias for the analysis presented here given our large
statistical error bars.  However, with the expected increase in the
size of the cluster catalogs from upcoming DES observations, it will
be necessary to constrain tSZ biases to better than 10\%.  There are
several possible ways to achieve this goal.  First, one can restrict
the cluster sample to even lower mass clusters, although this comes at
the cost of less signal-to-noise and reduced cosmological utility.
One can also apply more agressive filtering to remove the tSZ signal,
although this will also reduce the signal-to-noise.  Another option is
to attempt to model the tSZ signal or apply estimators which are
robust to its presence.  Finally, one can use the known frequency
dependence of the tSZ to construct multi-frequency combinations of CMB
maps for which the tSZ signal is minimized.  It is likely that this
last approach will prove essential for future CMB cluster lensing
analyses.

Some potential sources of contamination, however, cannot be eliminated
with multi-frequency information.  In particular, the kinematic SZ
(kSZ) signal imprinted on the CMB by clusters has the same frequency
dependence as the primordial CMB.  Because kSZ signal appears as a
monopole-like signal on the sky while the lensing signal is
dipole-like, it may be possible to effectively fit out the kSZ.
Alternatively, one may use polarization information to reconstruct the
lensing signal.  Because the polarized SZ signals are expected to be
much smaller than their temperature-only counterparts,
polarization-sensitive measurements (see below) offer a promising
route to obtaining unbiased estimates of the CMB cluster lensing
signal \citep[e.g.][]{Raghunathan:2017}.

To date, two methods have been applied to measure CMB lensing in the
one-halo regime: quadratic estimators (i.e.
\citealt{Madhavacheril:2015}, \citealt{Planck2015:XXIV} and this work)
and a maximum likelihood approach (i.e. \citetalias{Baxter:2015}).
While the maximum likelihood approach in principle offers higher
signal-to-noise, the quadratic estimator has the advantage that it is
quite robust to sources of contamination (such as tSZ) in the CMB
temperature maps\footnote{Note that this is not a fundamental
  limitation of maximum likelihood lensing mass estimation.  In
  principle, one could modify the maximum likelihood estimator to
  improve its robustness to sources of contamination.  However, the
  simple form of the maximum likelihood estimator considered in
  \citetalias{Baxter:2015} was found to be quite sensitive to various
  contaminants. }.  Indeed, the quadratic estimator approach was
employed here because it enabled the use of the 150 GHz SPT-SZ maps
despite the fact that these maps also have significant tSZ signal.
The 150 GHz maps have significantly higher signal-to-noise than the
tSZ-free linear combination of 90, 150 and 220 GHz maps generated for
the maximum likelihood analysis of \citetalias{Baxter:2015}.
Furthermore, as shown in \citet{Raghunathan:2017}, the increased
statistical power of the maximum likelihood estimator relative to the
quadratic estimator is small for SPT-SZ noise levels.

The reduced noise levels of future CMB experiments, however, make the
maximum likelihood estimator approach worth pursuing.  If low noise
levels can be achieved in tSZ-cleaned maps then maximum likelihood
cluster mass estimation may prove more powerful than the quadratic
estimator-based approach.  It may also be possible to modify the
maximum likelihood technique to increase its robustness to various
contaminants by e.g. applying additional filtering to the maps before
the estimator is applied.

The future of CMB cluster lensing with DES and SPT is exciting.  For
the Y1 cluster sample considered here, CMB cluster lensing is useful
primarily as a consistency check on the galaxy-lensing-inferred
cluster masses.  However, five-year DES observations will cover more
area and be significantly deeper than DES Y1 observations, resulting
in significantly expanded cluster samples, especially at high
redshifts.  As pointed out in \S\ref{sec:introduction}, it is at high
redshifts that CMB-cluster lensing has the potential to be competitive
with galaxy lensing.  Furthermore, new low-noise CMB experiments like
SPT-3G \citep{Benson:2014}, Advanced ACTpol \citep{Henderson:2016},
Simons Array \citep{Suzuki:2016}, Simons Observatory, and
CMB-S4 \citep{Abazajian:2016} are coming online soon that will
significantly improve the signal-to-noise of CMB cluster lensing
measurements \citep[e.g.][]{Louis:2017,Raghunathan:2017}.  These new
experiments will also provide low-noise measurements of the CMB
polarization signal, which as discussed above, will be useful for
constraining biases introduced by the SZ effect.

\section*{Acknowledgements}

This paper has gone through internal review by the DES collaboration.

EB is partially supported by the US Department of Energy grant
DE-SC0007901.  The Melbourne group acknowledges the support from
Australian Research Council's Discovery Projects scheme (DP150103208).
PF is funded by MINECO, projects ESP2013-48274-C3-1-P,
ESP2014-58384-C3-1-P, and ESP2015-66861-C3-1-R.  ER is supported by
DOE grant DE-SC0015975 and by the Sloan Foundation, grant FG-
2016-6443.

Funding for the DES Projects has been provided by the U.S. Department
of Energy, the U.S. National Science Foundation, the Ministry of
Science and Education of Spain, the Science and Technology Facilities
Council of the United Kingdom, the Higher Education Funding Council
for England, the National Center for Supercomputing Applications at
the University of Illinois at Urbana-Champaign, the Kavli Institute of
Cosmological Physics at the University of Chicago, the Center for
Cosmology and Astro-Particle Physics at the Ohio State University, the
Mitchell Institute for Fundamental Physics and Astronomy at Texas A\&M
University, Financiadora de Estudos e Projetos, Funda{\c c}{\~a}o
Carlos Chagas Filho de Amparo {\`a} Pesquisa do Estado do Rio de
Janeiro, Conselho Nacional de Desenvolvimento Cient{\'i}fico e
Tecnol{\'o}gico and the Minist{\'e}rio da Ci{\^e}ncia, Tecnologia e
Inova{\c c}{\~a}o, the Deutsche Forschungsgemeinschaft and the
Collaborating Institutions in the Dark Energy Survey.

The Collaborating Institutions are Argonne National Laboratory, the
University of California at Santa Cruz, the University of Cambridge,
Centro de Investigaciones Energ{\'e}ticas, Medioambientales y
Tecnol{\'o}gicas-Madrid, the University of Chicago, University College
London, the DES-Brazil Consortium, the University of Edinburgh, the
Eidgen{\"o}ssische Technische Hochschule (ETH) Z{\"u}rich, Fermi
National Accelerator Laboratory, the University of Illinois at
Urbana-Champaign, the Institut de Ci{\`e}ncies de l'Espai (IEEC/CSIC),
the Institut de F{\'i}sica d'Altes Energies, Lawrence Berkeley
National Laboratory, the Ludwig-Maximilians Universit{\"a}t
M{\"u}nchen and the associated Excellence Cluster Universe, the
University of Michigan, the National Optical Astronomy Observatory,
the University of Nottingham, The Ohio State University, the
University of Pennsylvania, the University of Portsmouth, SLAC
National Accelerator Laboratory, Stanford University, the University
of Sussex, Texas A\&M University, and the OzDES Membership Consortium.

Based in part on observations at Cerro Tololo Inter-American
Observatory, National Optical Astronomy Observatory, which is operated
by the Association of Universities for Research in Astronomy (AURA)
under a cooperative agreement with the National Science Foundation.

The DES data management system is supported by the National Science
Foundation under Grant Numbers AST-1138766 and AST-1536171.  The DES
participants from Spanish institutions are partially supported by
MINECO under grants AYA2015-71825, ESP2015-88861, FPA2015-68048,
SEV-2012-0234, SEV-2016-0597, and MDM-2015-0509, some of which include
ERDF funds from the European Union. IFAE is partially funded by the
CERCA program of the Generalitat de Catalunya.  Research leading to
these results has received funding from the European Research Council
under the European Union's Seventh Framework Program (FP7/2007-2013)
including ERC grant agreements 240672, 291329, and 306478.  We
acknowledge support from the Australian Research Council Centre of
Excellence for All-sky Astrophysics (CAASTRO), through project number
CE110001020.

This manuscript has been authored by Fermi Research Alliance, LLC
under Contract No. DE-AC02-07CH11359 with the U.S. Department of
Energy, Office of Science, Office of High Energy Physics. The United
States Government retains and the publisher, by accepting the article
for publication, acknowledges that the United States Government
retains a non-exclusive, paid-up, irrevocable, world-wide license to
publish or reproduce the published form of this manuscript, or allow
others to do so, for United States Government purposes.

The South Pole Telescope program is supported by the
National Science Foundation through grant PLR-1248097.
Partial support is also provided by the NSF Physics Frontier
Center grant PHY-0114422 to the Kavli Institute of Cosmological
Physics at the University of Chicago, the Kavli
Foundation, and the Gordon and Betty Moore Foundation
through Grant GBMF\#947 to the University of Chicago.
Argonne National Laboratory’s work was supported
under the U.S. Department of Energy contract DE-AC02-
06CH11357.

\bibliographystyle{mnras}
\bibliography{thebibliography}

\section*{Affiliations}

\textit{
$^{1}$ Department of Physics and Astronomy, University of Pennsylvania, Philadelphia, PA 19104, USA\\
$^{2}$ School of Physics, University of Melbourne, Parkville, VIC 3010, Australia\\
$^{3}$ Kavli Institute for Cosmological Physics, University of Chicago, Chicago, IL 60637, USA\\
$^{4}$ Department of Astronomy and Astrophysics, University of Chicago, Chicago, IL 60637, USA\\
$^{5}$ Institute of Space Sciences, IEEC-CSIC, Campus UAB, Carrer de Can Magrans, s/n,  08193 Barcelona, Spain\\
$^{6}$ Canadian Institute for Advanced Research, CIFAR Program on Gravity and the Extreme Universe, Toronto, ON, M5G 1Z8, Canada\\
$^{7}$ Astronomy Department, University of Illinois at Urbana Champaign, 1002 W. Green Street, Urbana, IL 61801, USA\\
$^{8}$ Department of Physics, University of Illinois Urbana Champaign, 1110 W. Green Street, Urbana, IL 61801, USA\\
$^{9}$ Department of Physics, McGill University, 3600 rue University, Montreal, QC, H3A 2T8, Canada\\
$^{10}$ Department of Physics, University of Arizona, Tucson, AZ 85721, USA\\
$^{11}$ Cerro Tololo Inter-American Observatory, National Optical Astronomy Observatory, Casilla 603, La Serena, Chile\\
$^{12}$ Fermi National Accelerator Laboratory, P. O. Box 500, Batavia, IL 60510, USA\\
$^{13}$ Department of Physics, University of California, Davis, CA 95616, USA\\
$^{14}$ CNRS, UMR 7095, Institut d'Astrophysique de Paris, F-75014, Paris, France\\
$^{15}$ Department of Physics \& Astronomy, University College London, Gower Street, London, WC1E 6BT, UK\\
$^{16}$ Sorbonne Universit\'es, UPMC Univ Paris 06, UMR 7095, Institut d'Astrophysique de Paris, F-75014, Paris, France\\
$^{17}$ Department of Physics, University of Chicago, 5640 South Ellis Avenue, Chicago, IL 60637, USA\\
$^{18}$ Argonne National Laboratory, 9700 South Cass Avenue, Lemont, IL 60439, USA\\
$^{19}$ Kavli Institute for Particle Astrophysics and Cosmology, Stanford University, 452 Lomita Mall, Stanford, CA 94305, USA\\
$^{20}$ SLAC National Accelerator Laboratory, Menlo Park, CA 94025, USA\\
$^{21}$ Enrico Fermi Institute, University of Chicago, Chicago, IL 60637, USA\\
$^{22}$ Laborat\'orio Interinstitucional de e-Astronomia - LIneA, Rua Gal. Jos\'e Cristino 77, Rio de Janeiro, RJ - 20921-400, Brazil\\
$^{23}$ Observat\'orio Nacional, Rua Gal. Jos\'e Cristino 77, Rio de Janeiro, RJ - 20921-400, Brazil\\
$^{24}$ National Center for Supercomputing Applications, 1205 West Clark St., Urbana, IL 61801, USA\\
$^{25}$ Institut de F\'{\i}sica d'Altes Energies (IFAE), The Barcelona Institute of Science and Technology, Campus UAB, 08193 Bellaterra (Barcelona) Spain\\
$^{26}$ California Institute of Technology, Pasadena, CA 91125, USA\\
$^{27}$ Department of Physics, University of California, Berkeley, CA 94720, USA\\
$^{28}$ Department of Physics, IIT Hyderabad, Kandi, Telangana 502285, India\\
$^{29}$ Faculty of Physics, Ludwig-Maximilians-Universit\"at, Scheinerstr. 1, 81679 Munich, Germany\\
$^{30}$ Excellence Cluster Universe, Boltzmannstr.\ 2, 85748 Garching, Germany\\
$^{31}$ Center for Astrophysics and Space Astronomy, Department of Astrophysical and Planetary Sciences, University of Colorado, Boulder, CO 80309, USA\\
$^{32}$ Instituto de Fisica Teorica UAM/CSIC, Universidad Autonoma de Madrid, 28049 Madrid, Spain\\
$^{33}$ European Southern Observatory, Karl-Schwarzschild Stra{\ss}e 2, 85748 Garching, Germany\\
$^{34}$ Institute of Astronomy, University of Cambridge, Madingley Road, Cambridge CB3 0HA, UK\\
$^{35}$ Kavli Institute for Cosmology, University of Cambridge, Madingley Road, Cambridge CB3 0HA, UK\\
$^{36}$ Universit\"ats-Sternwarte, Fakult\"at f\"ur Physik, Ludwig-Maximilians Universit\"at M\"unchen, Scheinerstr. 1, 81679 M\"unchen, Germany\\
$^{37}$ Department of Physics, University of Colorado, Boulder, CO 80309, USA\\
$^{38}$ Department of Physics, ETH Zurich, Wolfgang-Pauli-Strasse 16, CH-8093 Zurich, Switzerland\\
$^{39}$ Center for Cosmology and Astro-Particle Physics, The Ohio State University, Columbus, OH 43210, USA\\
$^{40}$ Department of Physics, The Ohio State University, Columbus, OH 43210, USA\\
$^{41}$ University of Chicago, Chicago, IL 60637, USA\\
$^{42}$ Astronomy Department, University of Washington, Box 351580, Seattle, WA 98195, USA\\
$^{43}$ Santa Cruz Institute for Particle Physics, Santa Cruz, CA 95064, USA\\
$^{44}$ Australian Astronomical Observatory, North Ryde, NSW 2113, Australia\\
$^{45}$ Physics Division, Lawrence Berkeley National Laboratory, Berkeley, CA 94720, USA\\
$^{46}$ Departamento de F\'isica Matem\'atica, Instituto de F\'isica, Universidade de S\~ao Paulo, CP 66318, S\~ao Paulo, SP, 05314-970, Brazil\\
$^{47}$ Steward Observatory, University of Arizona, 933 North Cherry Avenue, Tucson, AZ 85721, USA\\
$^{48}$ George P. and Cynthia Woods Mitchell Institute for Fundamental Physics and Astronomy, and Department of Physics and Astronomy, Texas A\&M University, College Station, TX 77843,  USA\\
$^{49}$ Department of Astronomy, The Ohio State University, Columbus, OH 43210, USA\\
$^{50}$ Department of Physics, University of Michigan, Ann Arbor, MI 48109, USA\\
$^{51}$ Department of Astrophysical Sciences, Princeton University, Peyton Hall, Princeton, NJ 08544, USA\\
$^{52}$ Department of Astronomy, University of Michigan, Ann Arbor, MI 48109, USA\\
$^{53}$ Instituci\'o Catalana de Recerca i Estudis Avan\c{c}ats, E-08010 Barcelona, Spain\\
$^{54}$ Max Planck Institute for Extraterrestrial Physics, Giessenbachstrasse, 85748 Garching, Germany\\
$^{55}$ Dunlap Institute for Astronomy \& Astrophysics, University of Toronto, 50 St George St, Toronto, ON, M5S 3H4, Canada\\
$^{56}$ Jet Propulsion Laboratory, California Institute of Technology, 4800 Oak Grove Dr., Pasadena, CA 91109, USA\\
$^{57}$ Department of Physics, University of Minnesota, Minneapolis, MN 55455, USA\\
$^{58}$ Center for Astrophysics and Space Astronomy, Department of Astrophysical and Planetary Sciences, University of Colorado, Boulder, CO, 80309\\
$^{59}$ NASA Postdoctoral Program Senior Fellow, NASA Ames Research Center, Moffett Field, CA 94035, USA\\
$^{60}$ Department of Physics and Astronomy, Pevensey Building, University of Sussex, Brighton, BN1 9QH, UK\\
$^{61}$ Physics Department, Center for Education and Research in Cosmology and Astrophysics, Case Western Reserve University,Cleveland, OH 44106, USA\\
$^{62}$ Centro de Investigaciones Energ\'eticas, Medioambientales y Tecnol\'ogicas (CIEMAT), Madrid, Spain\\
$^{63}$ School of Physics and Astronomy, University of Southampton,  Southampton, SO17 1BJ, UK\\
$^{64}$ Instituto de F\'isica Gleb Wataghin, Universidade Estadual de Campinas, 13083-859, Campinas, SP, Brazil\\
$^{65}$ Jet Propulsion Laboratory, California Institute of Technology, Pasadena, CA 91109, USA\\
$^{66}$ Harvard Smithsonian Center for Astrophysics, 60 Garden St, MS 12, Cambridge, MA, 02138, USA\\
$^{67}$ Dept. of Physics, Stanford University, 382 Via Pueblo Mall, Stanford, CA 94305, USA\\
$^{68}$ Computer Science and Mathematics Division, Oak Ridge National Laboratory, Oak Ridge, TN 37831, USA\\
$^{69}$ Institute of Cosmology \& Gravitation, University of Portsmouth, Portsmouth, PO1 3FX, UK\\
$^{70}$ Department of Astronomy \& Astrophysics, University of Toronto, 50 St George St, Toronto, ON, M5S 3H4, Canada\\
$^{71}$ Institute for Astronomy, University of Edinburgh, Edinburgh EH9 3HJ, UK\\
}

\bsp	
\label{lastpage}

\end{document}